# RIFTES: An RTM- and iteration-free temperature-emissivity separation framework for accurate and efficient clear-sky land surface temperature retrieval

Huanyu Zhang[a,c], Bo-Hui Tang[b,d,a,*], Yun Jiang[a], Menglin Si[a], Frank M. Göttsche[e], Tian Hu[f], Yuanliang Cheng[a,c], Zhao-Liang Li[a,d,g]

[a] *State Key Laboratory of Resources and Environmental Information System, Institute of Geographic Sciences and Natural Resources Research, Chinese Academy of Sciences, Beijing 100101, China*

[b] *Yunnan Key Laboratory of Quantitative Remote Sensing / Yunnan International Joint Laboratory for Integrated Sky-Ground Intelligent Monitoring of Mountain Hazards, Faculty of Land Resources Engineering, Kunming University of Science and Technology, Kunming 650093, China*

[c] *College of Resources and Environment, University of Chinese Academy of Sciences, Beijing 100049, China*

[d] *Southwest United Graduate School, Kunming 650092, China*

[e] *Institute of Meteorology and Climate Research, Karlsruhe Institute of Technology, Karlsruhe 76021, Germany*

[f] *Luxembourg Institute of Science and Technology, Belvaux 4362, Luxembourg*

[g] *State Key Laboratory of Efficient Utilization of Arid and Semi-arid Arable Land in Northern China, Institute of Agricultural Resources and Regional Planning, Chinese Academy of Agricultural Sciences, Beijing 100081, China*

[*] Corresponding author at: Faculty of Land Resource Engineering, Kunming University of Science and Technology, Kunming, 650093, China.

*E-mail address*: tangbh@kust.edu.cn (B.-H. Tang)

**Abstract:** The temperature-emissivity separation (TES) algorithm is one of the most widely used operational approaches for estimating clear-sky land surface temperature (LST), which is a fundamental parameter in thermal infrared remote sensing. Nevertheless, this method is limited by two major factors. First, the atmospheric correction step based on radiative transfer models (RTM) is computationally intensive, and its associated uncertainties may severely degrade the retrieval accuracy. Second, the iterative process involved in the normalized emissivity method (NEM) module further increases the computational burden. To address these challenges, this study proposes an RTM- and iteration-free TES (RIFTES) framework to improve both computational efficiency and retrieval accuracy. Based on physical derivations, a non-iterative TES algorithm was first developed by reformulating the original iterative procedure into a mathematically equivalent closed-form solution, thereby eliminating the need for cumbersome iterations. To further reduce error propagation risks and computational burdens, a deep residual neural network that integrates atmospheric radiative transfer physics was adopted to conduct atmospheric correction using easily accessible parameters, with a masking mechanism introduced to flexibly incorporate atmospheric constraints when available. Comprehensive validations demonstrate the effectiveness of the proposed algorithm. Simulation results show that RIFTES remains robust to input uncertainties and achieves the lowest root mean squared error (RMSE) of 1.06 K among representative existing algorithms, including split-window (SW), TES, and SW-TES hybrid methods. In-situ measurements from globally distributed sites were then used to evaluate the practical performance of RIFTES when applied to both the ECOsystem Spaceborne Thermal Radiometer Experiment on Space Station (ECOSTRESS) and the Advanced Baseline Imager (ABI). The new algorithm achieves RMSE values of 1.51 K and 1.97 K for ECOSTRESS and ABI, respectively, reducing retrieval uncertainties by up to 24% and 32% compared with existing methods. Furthermore, by simplifying both the iterative procedure and atmospheric correction, RIFTES reduces the overall computational time by 74.0% and 62.5% compared with the TES and hybrid algorithms, respectively. These results indicate that RIFTES not only improves retrieval accuracy and maintains stable cross-sensor performance, but also reduces input requirements and computational costs, demonstrating promising potential for operational applications.

# 1. Introduction

Land surface temperature (LST) is a key parameter in the Earth surface system, controlling the exchange of energy and water between the land surface and atmosphere (Li et al., 2023; Li et al., 2013). LST reflects the thermal state of the land surface, and is therefore of great importance for a wide range of applications, such as climate change studies (Lean, 2018; Wang et al., 2022; Zhou et al., 2022), drought monitoring (Anderson et al., 2016; Hu et al., 2020; Jia et al., 2025), surface radiation budget estimation (Liang et al., 2019; Liang et al., 2010), thermal anomaly detection (Bhardwaj et al., 2017; Blackett et al., 2011), and urban heat island analysis (Fu and Weng, 2016; Weng et al., 2004; Zhan et al., 2025).

Satellite remote sensing remains the most practical way for estimating LST at the global scale (Li et al., 2023), while the inherent coupling between LST and land surface emissivity (LSE) in remote sensing signals remains a key challenge. By introducing additional constraints derived from the relationship between spectral shape differences and minimum emissivity, the temperature-emissivity separation (TES) algorithm can simultaneously estimate LST and LSE from single-scene thermal infrared (TIR) observations (Gillespie et al., 1998). Such an advantage makes the TES algorithm one of the most widely used approaches for clear-sky LST retrieval, particularly for sensors operated by the National Aeronautics and Space Administration (NASA), such as the Advanced Spaceborne Thermal Emission and Reflection Radiometer (ASTER) (Gillespie et al., 1998), the Moderate Resolution Imaging Spectroradiometer (MODIS) (Hulley et al., 2018), the Visible Infrared Imaging Radiometer Suite (VIIRS) (Islam et al., 2017), and the ECOsystem Spaceborne Thermal Radiometer Experiment on Space Station (ECOSTRESS) (Hulley et al., 2022). Furthermore, the TES algorithm is also an important candidate for the next-generation TIR remote sensing missions (Cawse-Nicholson et al., 2021; Michel et al., 2021).

Despite its wide application, TES is computationally intensive, particularly compared with the split-window (SW) algorithm, another widely used operational approach. This high computational cost arises mainly from the preceding atmospheric correction step based on radiative transfer models (RTMs) and the complex iterative process involved in the normalized emissivity method (NEM) module. Furthermore, uncertainties associated with atmospheric correction may severely affect the retrieval accuracy of TES (Hulley et al., 2012; Zhang et al., 2025b), primarily due to errors in atmospheric profiles and the spatiotemporal interpolation required to match them with satellite observations (Ren et al., 2020; Zhang et al., 2025a).

Most attentions have been paid to refine the atmospheric correction step for TES. The water vapor scaling (WVS) method (Malakar and Hulley, 2016; Tonooka, 2001) is a classical approach for improving atmospheric correction accuracy (especially under humid conditions) but this improvement is achieved at the cost of substantially increased computational complexity due to repeated executions of RTMs and the TES algorithm (Liu et al., 2025; Zheng et al., 2024). In recent years, hybrid methods that incorporate the SW method into TES for atmospheric correction have emerged as the latest advances in clear-sky LST retrieval, which effectively simplify the atmospheric correction process and eliminate the reliance on atmospheric profiles (Ren et al., 2020; Wang et al., 2024; Wang et al., 2025; Zhang et al., 2025a; Zheng et al., 2022; Zheng et al., 2019). However, constrained by the limited modelling capacity of traditional SW methods, some approaches may yield lower retrieval accuracy than the conventional TES algorithm (Zhang et al., 2025a). Moreover, the validation of these emerging methods remains insufficient, particularly in terms of comparisons with representative algorithms and cross-sensor performance assessments based on comprehensive in-situ measurements and simulation datasets.

In contrast, the iterative procedure within the TES algorithm has received relatively limited attention. The NEM module requires up to a dozen iterations for each execution, and our experimental results show that its execution time is generally longer than that of the atmospheric correction step. As a result, simplifying atmospheric correction alone, as in existing hybrid methods, may be insufficient to substantially improve the overall computational efficiency. However, to the best of our knowledge, no existing study has attempted to facilitate the efficient implementation of TES or TES-based hybrid methods by simplifying the iterative procedure.

Therefore, despite extensive efforts to refine the TES algorithm, existing methods still struggle to simultaneously achieve high retrieval accuracy and high computational efficiency. Furthermore, a comprehensive benchmark validation comparing the performance of conventional methods (e.g., SW and TES) with emerging hybrid methods is still lacking, limiting a systematic understanding of the strengths and limitations of various algorithms as well as the operational potential of hybrid methods.

To address these research gaps, this study proposed a novel RTM- and iteration-free TES (RIFTES) framework for accurate and efficient clear-sky LST retrieval.

Specifically, based on physical derivations, the iterative procedure in the NEM module was reformulated into a mathematically equivalent closed-form solution that directly yields the converged result of the original iterative process, thereby eliminating cumbersome iterations and boosting computational efficiency. To further reduce the error propagation risk and computational burden, a deep neural network was adopted to replace RTMs for atmospheric correction, which was improved from our previous study (Zhang et al., 2025a) by introducing a new masking mechanism to flexibly incorporate atmospheric constraints when available. Simulation datasets covering diverse global atmospheric and surface conditions, together with in-situ measurements from globally distributed sites, were used to provide a comprehensive benchmark validation of RIFTES and other representative methods across different sensors. Finally, the computational efficiency of different algorithms was also fairly compared.

The remainder of this paper is organized as follows. Section 2 introduces the data used in this study. Section 3 describes the basic principles of the RIFTES framework and evaluation strategies in detail. The results and discussion are presented in Sections 4 and 5, respectively. Finally, Section 6 provides a brief conclusion.

## 2. Data

Two representative sensors were considered in this study to evaluate the cross-sensor performance of different algorithms. The first is NASA's ECOsystem Spaceborne Thermal Radiometer Experiment on Space Station (ECOSTRESS) (Fisher et al., 2020), which serves as a precursor to next-generation TIR missions and provides high-spatial-resolution (~70 m) observations with a revisiting interval of 1–3 days. The second is the Advanced Baseline Imager (ABI) onboard the Geostationary Operational Environmental Satellite-16 (GOES-16) operated by the National Oceanic and Atmospheric Administration (NOAA), which provides high-temporal-resolution observations every 10 minutes at a relatively coarse spatial resolution of ~2 km. The datasets used for algorithm application and validation across these two sensors are described in the following sections.

### 2.1 The simulation dataset

Simulation datasets generated through radiative transfer processes are particularly suitable for model training and evaluation, as they can capture diverse atmospheric and surface conditions worldwide in a cost-effective manner while implicitly embedding atmospheric radiative transfer physics into models during training.

Eq. (1) provides the simplified atmospheric radiative transfer equation under the assumption of a homogeneous land surface and a horizontally uniform atmosphere:

$$L_i^{TOA}(\theta_v) = \tau_i(\theta_v)L_i^{grd} + L_i^{atm\uparrow}(\theta_v) \tag{1}$$

where the subscript $i$ denotes the $i$th band of the sensor, $L_i^{TOA}$ represents TOA radiance, $\theta_v$ is the view zenith angle (VZA), $\tau_i$ is atmospheric transmittance, $L_i^{grd}$ represents ground-leaving radiance, and $L_i^{atm\uparrow}$ is atmospheric upward radiance. $L_i^{grd}$ can be calculated as follows:

$$L_i^{grd} = \varepsilon_i B_i(T_s) + (1 - \varepsilon_i)L_i^{atm\downarrow} \tag{2}$$

where $\varepsilon_i$ represent channel-effective LSE, $B_i(\cdot)$ denotes the Planck function, $T_s$ is LST, and $L_i^{atm\downarrow}$ is the hemispherical averaged atmospheric downward radiance. $\tau_i$, $L_i^{atm\uparrow}$, and $L_i^{atm\downarrow}$ are collectively referred to as the atmospheric parameters. In the following text, vectors composed of the values of a given variable across all bands are denoted in bold.

Based on Eqs. (1) and (2), our previous study (Zhang et al., 2025a) constructed a simulation dataset for ECOSTRESS using the recently released global atmospheric profile database (Ermida and Trigo, 2022) and the ECOSTRESS spectral library (Baldridge et al., 2009; Meerdink et al., 2019). Specifically, atmospheric parameters were estimated by inputting atmospheric profiles into the Radiative Transfer for TOVS (RTTOV) (Saunders et al., 2018), while possible LST and LSE were obtained from the atmospheric profile database and spectral library, respectively. The simulation dataset included 10,144,350 samples in total (67,629 groups of atmospheric parameters * 6 random VZAs * 5 possible LSTs * 5 possible LSEs), which was directly used in this study for model training and simulation analysis for the ECOSTRESS sensor.

Following the same data sources and the similar methodology, this study further compiled another simulation dataset for the ABI sensor. Specifically, 15,089 profiles spanning 130°W–30°W and 60°S–65°N were selected, covering most of the GOES-16 ABI full disk. For each profile, 5 random VZAs in each group (i.e., [0°, 15°], [15°, 30°], [30°, 45°], [45°, 60°], and [60°, 75°]) were generated, 5 possible LSTs calibrated by remote sensing data were used, and 5 LSEs in the spectral library were randomly assigned according to the land cover type. Finally, a representative simulation dataset including 1,886,125 samples (15,089 groups of atmospheric parameters * 5 random VZAs * 5 possible LSTs * 5 possible LSEs) was compiled for GOES-16 ABI, thereby facilitating the transfer learning of the neural network.

## 2.2 Remote sensing and reanalysis data

For ECOSTRESS, the latest Collection 2 (C2) products between 2019 and 2025 were downloaded, including ECO_L1B_RAD v002 containing TOA radiances for the five TIR bands, ECO_L1B_GEO v002 for geolocation and view/solar geometry, ECO_L2_LSTE v002 for LST and precipitable water vapor (PWV) information, and ECO_L2_CLOUD v002 for cloud masking. The new C2 product incorporates important improvements in radiometric calibration and cloud masking (Zhang et al., 2025c). As observations from ECOSTRESS Bands 1 and 3 were unavailable for an extended period from 2019 to 2023 due to a failure of the mass storage unit (Hulley et al., 2022), TOA radiances from the other three bands (i.e., ECOSTRESS Bands 2, 4, and 5) were used in this study.

For GOES-16 ABI, the Cloud and Moisture Imagery product (CMIP) and LST product released by NOAA at each hourly timestamp in 2020 were used. CMIP provides TOA reflectance factor for reflective bands and brightness temperature data for emissive bands. In this study, TOA brightness temperature from ABI Bands 11, 13, and 15 were used, as their spectral configurations are similar to those of the selected ECOSTRESS bands (Fig. 1), which facilitates the transfer learning of the neural network. The brightness temperature can be easily converted to radiance using the Planck function. The official LST product is generated using the SW algorithm (Yu et al., 2009), with cloud mask information provided alongside the product. To apply the RIFTES and hybrid algorithms, PWV information aligned with the ABI observation times was extracted from the ERA5 reanalysis product. To reproduce the conventional TES algorithm, atmospheric temperature and humidity profiles at 37 pressure levels, along with surface skin temperature, 2 m air temperature, and surface pressure at the single level, were extracted from the ERA5 reanalysis data to drive RTTOV. All reanalysis data were bilinearly interpolated to the 2-km grids.

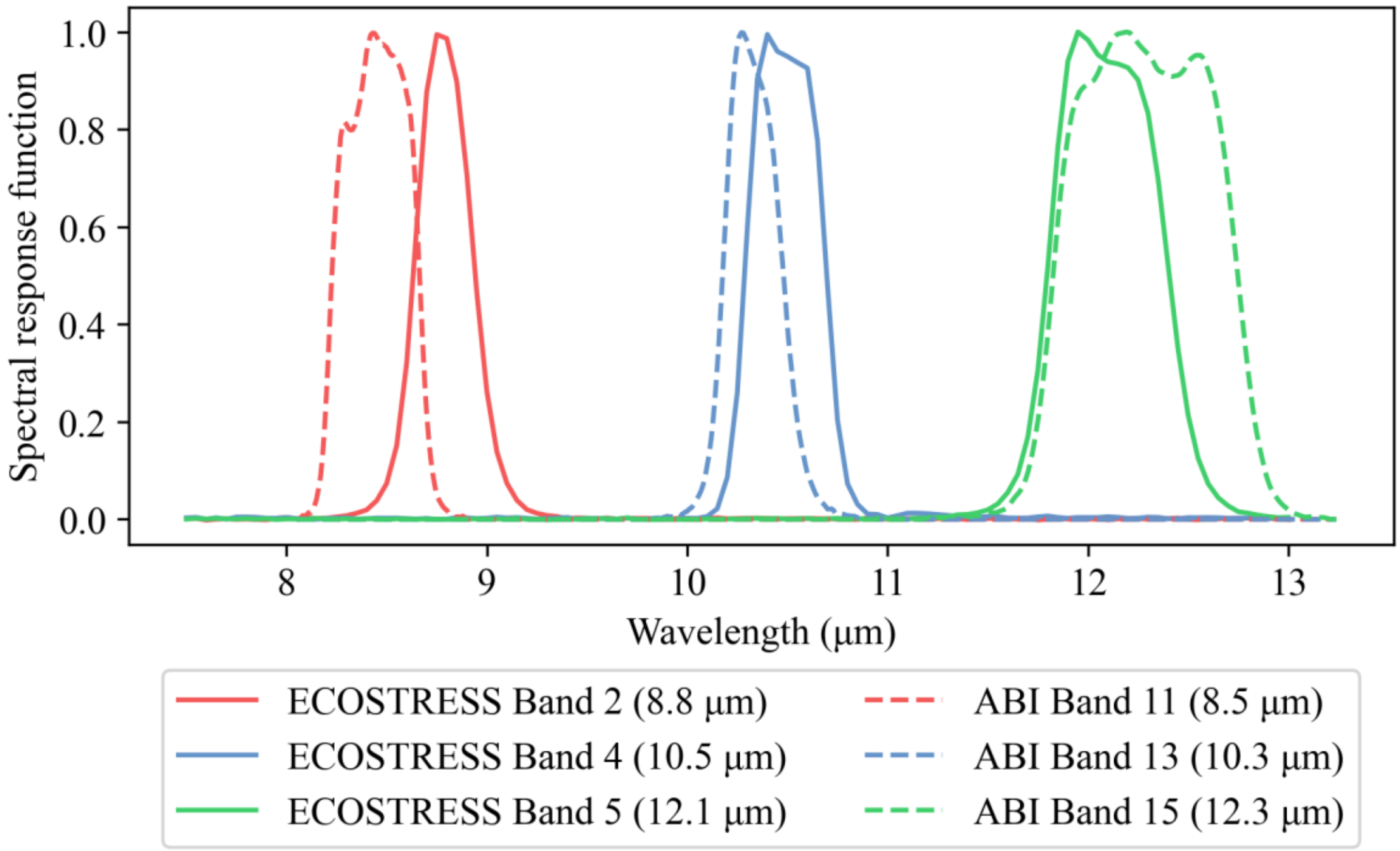

**Fig. 1.** The utilized thermal infrared bands of ECOSTRESS and ABI in this study.

### 2.3 In-situ measurements

In-situ measurements coincident with satellite overpasses were used for validating the practical accuracy of satellite-derived LST estimates. For ECOSTRESS, five radiometer sites located in Europe and Africa were used. The first two sites are from the Karlsruhe Institute of Technology (KIT) network (Göttsche et al., 2016; Pérez-Planells and Göttsche, 2023), namely CNS in the Lake Constance region between Germany, Switzerland, and Austria, and GBBp on the gravel plains near Gobabeb, Namibia. The forest site KIT from the Copernicus LAW network (Pérez-Planells and Göttsche, 2023) was also selected. Finally, two sites from the Global Change Unit (GCU) network in Spain (Sobrino and Skoković, 2016) were selected, with one located in shrublands (CDG) and the other in wetlands (FDU). For GOES-16 ABI, all seven pyrgeometer sites from the Surface Radiation Budget Network (SURFRAD) (Augustine et al., 2000) were used, namely BON, DRA, FPK, GWN, PSU, SXF, and TBL, which are distributed across different regions of the United States. Fig. 2 shows the spatial distributions of these radiometer and pyrgeometer sites, with the corresponding detailed information given in Table 1. These sites are located over various land cover types and climate zones, ensuing a relatively comprehensive validation result.

In-situ LST from sites specifically set up for LST validation and equipped with radiometers generally provide higher validation confidence and agree better with satellite-derived LST (Hu et al., 2022; Trigo et al., 2021; Zhang et al., 2025a). In contrast, validation results based on pyrgeometer sites are typically more affected by factors such as site heterogeneity, thermal directionality, and shortwave radiation contamination, resulting in slightly poorer consistency with satellite-derived LST. Nevertheless, pyrgeometer sites from the SURFRAD network remain widely used for LST validation (Duan et al., 2019; Liu et al., 2025; Sun and Pinker, 2003; Trigo et al., 2021; Wang et al., 2025; Yu et al., 2012), as they are generally well maintained and the temporal resolution is high (i.e., 1 minute). Therefore, this study finally selected these sites for T-based validation.

For radiometer sites, the corresponding in-situ LST is obtained using the Planck function:

$$T_s = B_k^{-1}\left[\frac{L_k^{\uparrow} - (1-\varepsilon_k)L_k^{\downarrow}}{\varepsilon_k}\right] \tag{3}$$

where the subscript $k$ denotes the narrow band of the radiometer (e.g., typically 9.6–11.5 μm or 8–14 μm), $B_k^{-1}$ is the Inverse Planck function, $L_k^{\uparrow}$ and $L_k^{\downarrow}$ are band-effective upward and downward radiances, and $\varepsilon_k$ is the band-effective LSE. $\varepsilon_k$ is generally determined by lab or in-situ measurements (Göttsche et al., 2016; Sobrino and Skoković, 2016).

For pyrgeometer sites, the in-situ LST can be calculated from the Stefan-Boltzmann law:

$$T_s = \sqrt[4]{\frac{R^{\uparrow} - (1 - \varepsilon_{BB})R^{\downarrow}}{\varepsilon_{BB}\sigma}} \tag{4}$$

where $R^{\uparrow}$ and $R^{\downarrow}$ represent the upward and downward longwave broadband radiations (e.g., typically about 4–50 μm), $\varepsilon_{BB}$ is the broadband emissivity, and $\sigma$ represents the Stefan-Boltzmann constant ($5.67\times10^{-8}$ W m$^{-2}$ K$^{-4}$). In this study, $\varepsilon_{BB}$ is estimated from the ASTER Global Emissivity Dataset (GED) v003 product using the linear conversion relationship proposed by Cheng et al. (2013).

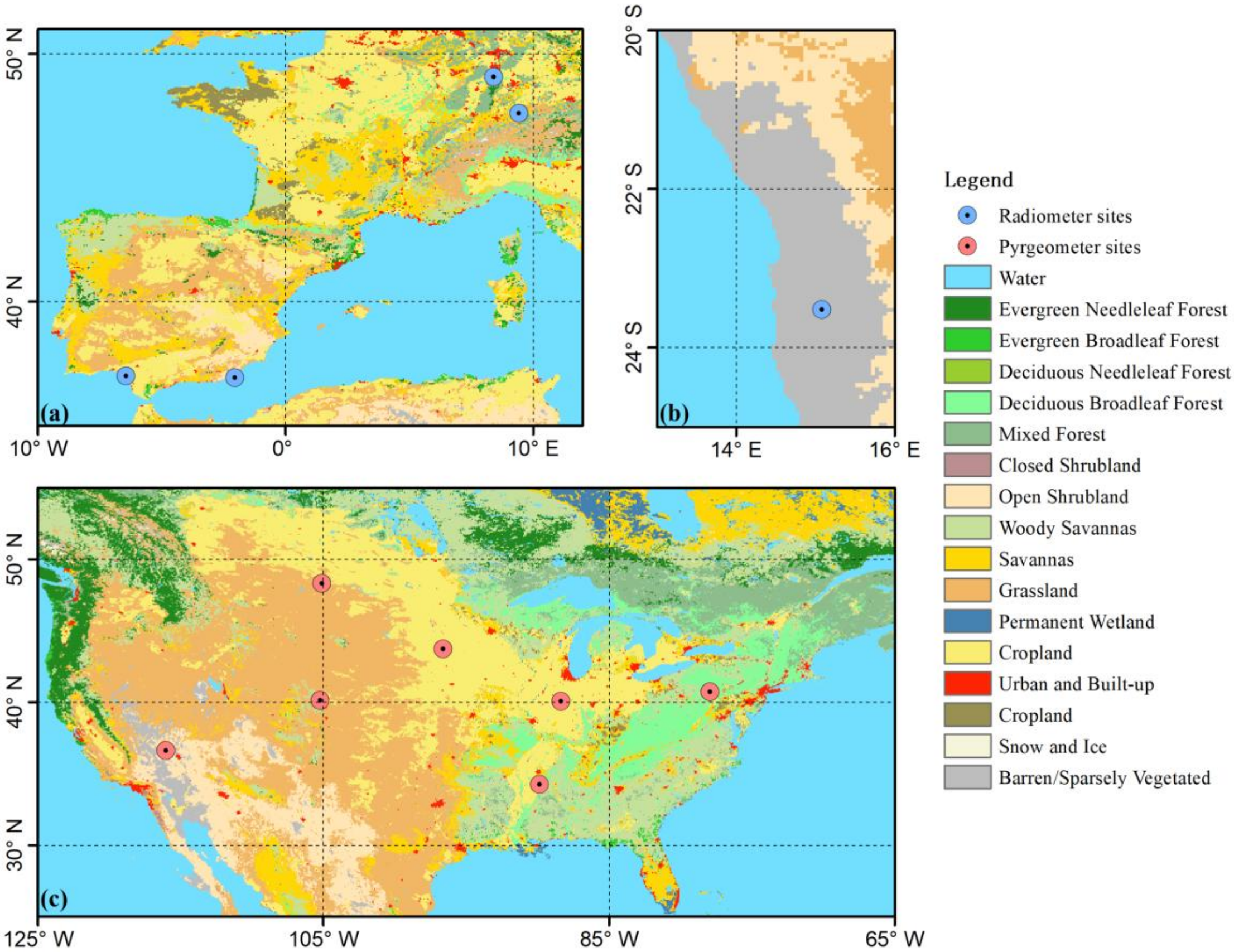

**Fig. 2.** Spatial distribution of selected validation sites. (a) Sites CNS, KIT, CDG, and FDU in Europe, (b) Site GBBp in Namibia, (c) SURFRAD sites in United States.

**Table 1.** Basic information of the selected sites for T-based validation.

| Site | Network | Latitude | Longitude | Land cover |
|---|---|---|---|---|
| CNS | KIT | 47.605°N | 9.444°E | Water |
| GBBp | KIT | 23.520°S | 15.083°E | Barren |
| KIT | Copernicus | 49.091°N | 8.425°E | Forest |
| CDG | GCU | 36.939°N | 2.034°W | Shrubland |
| FDU | GCU | 36.998°N | 6.434°W | Wetland |
| BON | SURFRAD | 40.052°N | 88.373°W | Cropland |
| DRA | SURFRAD | 36.624°N | 116.019°W | Barren |
| FPK | SURFRAD | 48.308°N | 105.102°W | Grassland |
| GWN | SURFRAD | 34.255°N | 89.873°W | Grassland |
| PSU | SURFRAD | 40.720°N | 77.931°W | Cropland |
| SXF | SURFRAD | 43.734°N | 96.623°W | Grassland |
| TBL | SURFRAD | 40.125°N | 105.237°W | Grassland |

# 3. Methods

## 3.1 Non-iterative TES algorithm

### 3.1.1 The original NEM module

The original NEM module iteratively removes $\boldsymbol{L}^{atm\downarrow}$ from $\boldsymbol{L}^{grd}$ to eliminate the atmospheric effect. Before iteration, the initial emissivity (i.e., $\varepsilon_i^0$) for each band is set to the possible maximum value (e.g., 0.99 for gray bodies and 0.96 for non-gray bodies). Then, for the arbitrary *i*-th band of the sensor during the first iteration, its corresponding LST is calculated from Eq. (2) as follows:

$$T_{s,i}^1 = B_i^{-1}\left[\frac{L_i^{grd} - (1-\varepsilon_i^0)L_i^{atm\downarrow}}{\varepsilon_i^0}\right] \tag{5}$$

where the superscript number denotes the iteration round. In this study, *i* takes values of 1, 2, and 3, corresponding to ECOSTRESS Bands 2, 4, and 5, or GOES-16 ABI Bands 11, 13, and 15, respectively.

In the first iteration, the maximum LST value across all bands is assigned as the NEM LST:

$$T_{s,NEM}^1 = \max(\boldsymbol{T_s^1}) \tag{6}$$

where $T_{s,NEM}^1$ represents the NEM LST after first iteration, and $\max(\cdot)$ denotes the maximum operation applied to a vector.

After obtaining $T_{s,NEM}^1$, the emissivity value for the *i*-th band is updated as follows:

$$\varepsilon_i^1 = \frac{L_i^{grd} - (1-\varepsilon_i^0)L_i^{atm\downarrow}}{B_i\left(T_{s,NEM}^1\right)} \tag{7}$$

Similarly, in the second iteration, the LST and LSE of the *i*-th band are calculated as follows:

$$\begin{cases} T_{s,i}^2 = B_i^{-1}\left[\dfrac{L_i^{grd} - (1-\varepsilon_i^1)L_i^{atm\downarrow}}{\varepsilon_i^1}\right] \\ T_{s,NEM}^2 = \max(\boldsymbol{T_s^2}) \\ \varepsilon_i^2 = \dfrac{L_i^{grd} - (1-\varepsilon_i^1)L_i^{atm\downarrow}}{B_i\left(T_{s,NEM}^2\right)} \end{cases} \tag{8}$$

The iterations continue until convergence, divergence, unreasonable emissivity values are detected, or the maximum number of iterations (e.g., 12–15) is reached. If convergence is achieved, the LST and LSE obtained in the final iteration are output as the NEM LST and LSE. Otherwise, the NEM iteration is regarded as unsuccessful, and the corresponding quality assurance (QA) flag is reported. In practical applications, the NEM module usually requires multiple iterations, which can be time-consuming. Therefore, eliminating the complex iterative procedure would greatly improve the overall computational efficiency of the TES algorithm.

### 3.1.2 The non-iterative NEM module

In the first iteration of the original NEM module for the *i*-th band, since $B_i\left(T_{s,NEM}^1\right) \geq B_i\left(T_{s,i}^1\right)$, it follows that:

$$\varepsilon_i^1 = \frac{L_i^{grd} - (1 - \varepsilon_i^0)L_i^{atm\downarrow}}{B_i\left(T_{s,NEM}^1\right)} \leq \frac{L_i^{grd} - (1 - \varepsilon_i^0)L_i^{atm\downarrow}}{B_i\left(T_{s,i}^1\right)} = \varepsilon_i^0 \tag{9}$$

Eq. (9) indicates that $\varepsilon_i^1 \leq \varepsilon_i^0$. Therefore, the following relationship can be derived:

$$\varepsilon_i^1 = \frac{L_i^{grd} - (1 - \varepsilon_i^0)L_i^{atm\downarrow}}{B_i\left(T_{s,NEM}^1\right)} \geq \frac{L_i^{grd} - (1 - \varepsilon_i^1)L_i^{atm\downarrow}}{B_i\left(T_{s,NEM}^1\right)} \tag{10}$$

By rearranging Eq. (5), $\varepsilon_i^0$ can be expressed as follows:

$$\varepsilon_i^0 = \frac{L_i^{grd} - L_i^{atm\downarrow}}{B_i\left(T_{s,i}^1\right) - L_i^{atm\downarrow}} \tag{11}$$

As $\varepsilon_i^0$ is positive, if $L_i^{grd} > L_i^{atm\downarrow}$, which is the most common case, then $B_i\left(T_{s,i}^1\right) > L_i^{atm\downarrow}$. Conversely, for the rare case where $L_i^{grd} < L_i^{atm\downarrow}$, $B_i\left(T_{s,i}^1\right) < L_i^{atm\downarrow}$. Therefore, Eq. (10) can be rearranged as follows:

$$\begin{cases} \varepsilon_i^1 \geq \dfrac{L_i^{grd} - L_i^{atm\downarrow}}{B_i\left(T_{s,NEM}^1\right) - L_i^{atm\downarrow}} & \text{when } L_i^{grd} > L_i^{atm\downarrow} \\ \varepsilon_i^1 \leq \dfrac{L_i^{grd} - L_i^{atm\downarrow}}{B_i\left(T_{s,NEM}^1\right) - L_i^{atm\downarrow}} & \text{when } L_i^{grd} < L_i^{atm\downarrow} \end{cases} \tag{12}$$

From Eqs. (8) and (12), LST of the *i*-th band in the second iteration follows:

$$\begin{cases} B_i\left(T_{s,i}^2\right) = L_i^{atm\downarrow} + \dfrac{L_i^{grd} - L_i^{atm\downarrow}}{\varepsilon_i^1} \leq B_i\left(T_{s,NEM}^1\right) & \text{when } L_i^{grd} > L_i^{atm\downarrow} \\ B_i\left(T_{s,i}^2\right) = L_i^{atm\downarrow} - \dfrac{L_i^{grd} - L_i^{atm\downarrow}}{\varepsilon_i^1} \leq B_i\left(T_{s,NEM}^1\right) & \text{when } L_i^{grd} < L_i^{atm\downarrow} \end{cases} \tag{13}$$

Therefore, regardless of the relative magnitudes of $L_i^{grd}$ and $L_i^{atm\downarrow}$, the LST of arbitrary band in the second round (i.e., $T_{s,i}^2$) does not exceed $T_{s,NEM}^1$, and consequently, the NEM LST remains unchanged. Similarly, it is easy to know that the NEM LST will not change in the subsequent iterations. In other words, the NEM LST is already determined after the first iteration. Given this phenomenon, the NEM LSE (i.e., $\boldsymbol{\varepsilon_{NEM}}$) can be directly derived from the maximum LST after the first iteration, forming the basis of the new non-iterative NEM module:

$$\begin{cases} B_i\left(T_{s,i}^1\right) = \dfrac{L_i^{grd} - (1 - \varepsilon_i^0)L_i^{atm\downarrow}}{\varepsilon_i^0} \\ T_{s,NEM}^1 = \max(\boldsymbol{T_s^1}) \\ \boldsymbol{\varepsilon_{NEM}} = \dfrac{\boldsymbol{L^{grd}} - \boldsymbol{L^{atm\downarrow}}}{\boldsymbol{B}\left(T_{s,NEM}^1\right) - \boldsymbol{L^{atm\downarrow}}} \end{cases} \quad (14)$$

In theory, $\boldsymbol{\varepsilon_{NEM}}$ from the non-iterative NEM module is equal to the asymptotic value of LSE estimates from the original NEM module (denoted as $\boldsymbol{\tilde{\varepsilon}_{NEM}}$) after infinite iterations:

$$\boldsymbol{\varepsilon_{NEM}} = \lim_{n\to\infty} \boldsymbol{\tilde{\varepsilon}_{NEM}}(n) \quad (15)$$

According to the above physical derivations and analyses, the iterative procedure in the original NEM module can be reformulated as a mathematically equivalent closed-form solution, which is expected to directly provide the converged result and therefore significantly boost computational efficiency. However, it should be noted that certain checks performed during the iterative process are no longer applicable. The first is the check on whether the number of iterations exceeds a predefined threshold, which becomes unnecessary since the outputs from the non-iterative NEM module are theoretically equivalent to the asymptotic values of the original NEM module after infinite iterations. The divergence test is likewise no longer applicable; nevertheless, such cases rarely occur and were not observed in the following simulations or practical applications. Therefore, the new non-iterative NEM module determines the validity of the results based on whether the estimated LSE values fall within a physically reasonable range.

### 3.1.3 The RATIO and MMD modules

Both the conventional and non-iterative NEM modules provide initial estimates of LST and LSE. Owing to uncertainties in the predefined maximum LSE values, the NEM LST and LSE are inevitably subject to errors (Zhang et al., 2025b). Therefore, the RATIO module was applied to estimate emissivity ratios from the NEM LSE, as these ratios are less sensitive to errors in LST estimates than absolute emissivity values (Watson, 1992):

$$\boldsymbol{\beta} = \frac{\boldsymbol{\varepsilon_{NEM}}}{\mathrm{mean}(\boldsymbol{\varepsilon_{NEM}})} \quad (16)$$

where $\boldsymbol{\beta}$ represents emissivity ratios and $\text{mean}(\cdot)$ denotes the mean operation applied to a vector.

After obtaining the relatively accurate emissivity ratios, the MMD module was applied to retrieve emissivity magnitudes from an empirical power-law relationship:

$$\begin{cases} \text{MMD} = \max(\boldsymbol{\beta}) - \min(\boldsymbol{\beta}) \\ \varepsilon_{\min} = a - b \times \text{MMD}^{c} \\ \boldsymbol{\varepsilon} = \dfrac{\varepsilon_{\min}}{\min(\boldsymbol{\beta})}\boldsymbol{\beta} \end{cases} \tag{17}$$

where $\min(\cdot)$ denotes the minimum operation applied to a vector, $a$, $b$, and $c$ are coefficients of the empirical relationship. For ECOSTRESS, $a$=0.9895, $b$=0.7994, and $c$=0.8572. For GOES-16, $a$=0.9872, $b$=0.7781, $c$=0.8434.

Finally, LST was recalculated from the band with the maximum emissivity to reduce the uncertainty of atmospheric correction:

$$T_s = B_*^{-1}\left[\frac{L_*^{grd} - (1 - \varepsilon_*)L_*^{atm\downarrow}}{\varepsilon_*}\right] \tag{18}$$

where the subscript * denote the band with the maximum LSE value.

The non-iterative NEM module, together with the RATIO and MMD modules, forms the new non-iterative algorithm, whereas the conventional NEM module combined with the same two modules constitutes the conventional TES algorithm.

## 3.2 RTM-free atmospheric correction model

The non-iterative TES algorithm requires atmospherically corrected radiances as inputs. For TIR atmospheric correction, neural networks have proven capable of reducing input requirements and computational time compared with RTMs, which also exhibit stronger modeling capability than traditional SW equations (Zhang et al., 2025a). Therefore, building on the original model structure (Zhang et al., 2025a), this study further proposed a deep residual neural network with masking mechanism (DRNNMM) for atmospheric correction, which can operate with or without PWV input using readily accessible satellite observations.

### 3.2.1 Model structure

Fig. 3 illustrates the detailed structure of DRNNMM, which is composed of a masking layer, an input projection layer, several sequential residual blocks (ResBlocks) as the backbone, and an output layer. The input parameters include TOA radiances, VZA, PWV, and a binary mask.

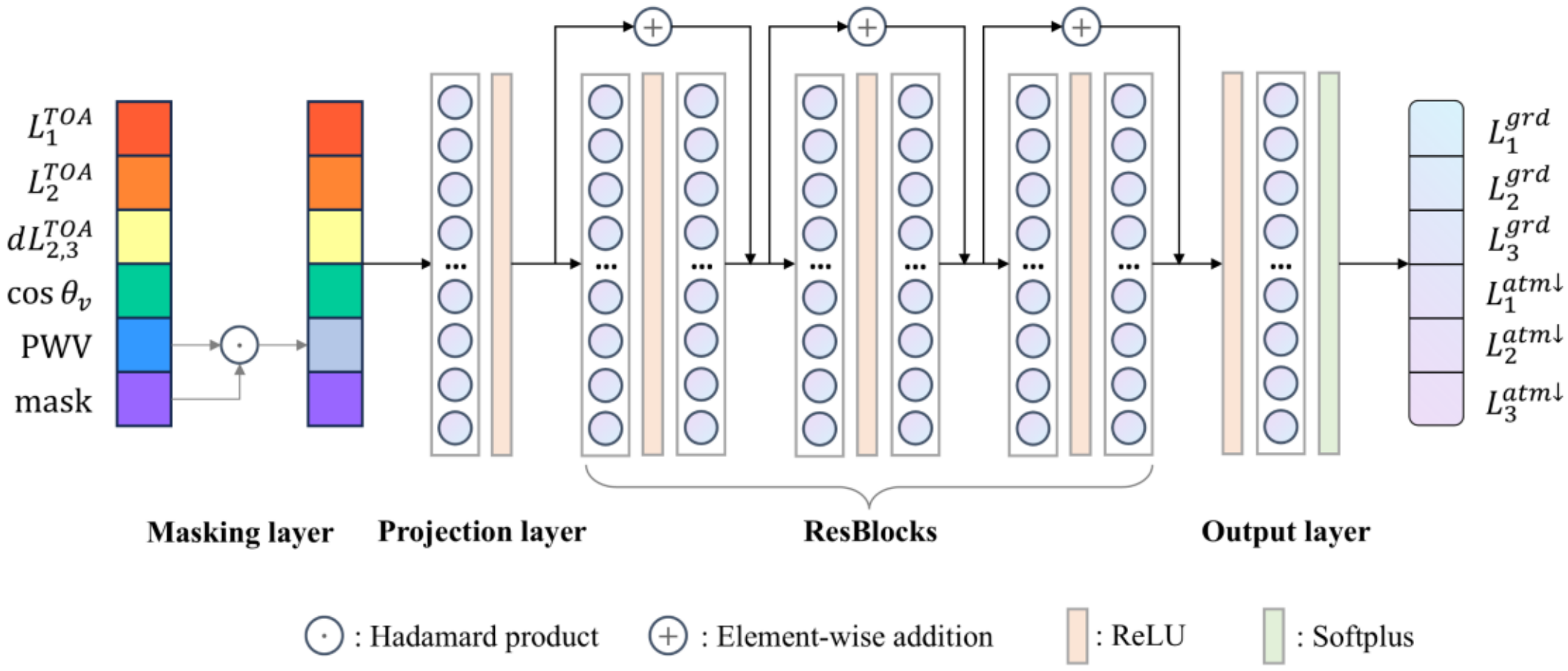

**Fig. 3.** Detailed structure of the DRNNMM.

The masking layer selectively applies a mask to PWV depending on its availability, which allows the model to flexibly accommodate both the presence and absence of PWV input. The basic principle of this layer is summarized as follows:

$$\begin{cases} \mathrm{PWV_{masked}} = \mathrm{PWV} \odot \mathrm{mask} \\ \mathrm{mask} \in \{0, 1\} \end{cases} \tag{19}$$

where $\odot$ denotes the Hadamard product.

Then, the input projection layer was used to extract shallow features from the masked vector, mapping the 6-dimensional input to a 128-dimensional hidden representation. Subsequently, the features passed sequential ResBlocks for extracting the deep features about surface and atmospheric characteristics. Finally, the output layer was applied to estimate $\boldsymbol{L^{grd}}$ and $\boldsymbol{L^{atm\downarrow}}$ from the 128-dimensional high-level features. A Softplus activation function layer was used in the end to enforce non-negative radiance estimates to prevent physically unrealistic solutions, although negative solutions rarely occur in practice.

The overall prediction process of DRNNMM can be abstracted as the following implicit function:

$$\hat{\boldsymbol{L}}^{\boldsymbol{grd}}, \hat{\boldsymbol{L}}^{\boldsymbol{atm\downarrow}} = \mathrm{DRNNMM}\left(L_1^{TOA}, L_2^{TOA}, dL_{2,3}^{TOA}, \cos\theta_v, \mathrm{PWV}, \mathrm{mask}\right) \tag{20}$$

where $L_1^{TOA}$ represents the TOA radiance from the band located within 8–10 μm, $L_2^{TOA}$ and $L_3^{TOA}$ are TOA radiances of two bands within 10–13 μm, the difference $dL_{2,3}^{TOA} = L_3^{TOA} - L_2^{TOA}$ reflects the fundamental principle of the SW method which eliminates atmospheric effects by exploiting the differences between two adjacent TIR bands, $\cos\theta_v$ reflects the slant atmospheric path and its associated atmospheric effects, PWV is also a common indicator of atmospheric effect, and the binary mask indicates whether

PWV is used.

Therefore, DRNNMM is capable of estimating $\boldsymbol{L^{grd}}$ and $\boldsymbol{L^{atm\downarrow}}$ from readily accessible satellite observations and optionally available PWV information. Compared to conventional atmospheric correction methods that rely on atmospheric RTMs and complete atmospheric profiles, DRNNMM effectively simplifies the input parameters and radiative transfer calculations, which is expected to simultaneously reduce the risk of error propagation and computational complexity.

### 3.2.2 Model training and transferring

DRNNMM was first trained on the simulation dataset compiled for ECOSTRESS, with parameters updated based on the estimation errors of radiances. According to our previous experimental results, different weightings between $\boldsymbol{L^{grd}}$ and $\boldsymbol{L^{atm\downarrow}}$ have only a minor impact on the final accuracy (Zhang et al., 2025a). Therefore, for simplicity, an equal-weighting scheme is employed in this study, and the final loss function (i.e., mean squared error of radiances) is written as follows:

$$\mathcal{L} = \left\| \left[ \boldsymbol{L_{true}^{grd}}, \boldsymbol{L_{true}^{atm\downarrow}} \right] - \left[ \boldsymbol{\hat{L}^{grd}}, \boldsymbol{\hat{L}^{atm\downarrow}} \right] \right\|^2 \tag{21}$$

where $\boldsymbol{L_{true}^{grd}}$ and $\boldsymbol{L_{true}^{atm\downarrow}}$ denote the true values of radiances, respectively, and the operator $[\cdot, \cdot]$ denotes vector concatenation.

During training, the binary mask was randomly generated with equal probability (50%) of being 0 or 1, enabling the model to flexibly adapt to scenarios with or without PWV input. The number of epochs was set to 200 and the batch size was set to 2048. The AdamW optimizer was adopted with an initial learning rate of $10^{-3}$ and a weight decay of $10^{-2}$. The learning rate was reduced by a factor of 10 if the loss on the validation set did not decrease for 5 consecutive epochs, with a minimum learning rate setting to $10^{-5}$. Early stopping strategy was applied if the validation loss did not decrease for 10 consecutive epochs.

The DRNNMM model trained for ECOSTRESS can be readily adapted to other sensors through transfer learning using sensor-specific simulation datasets. Due to discrepancies in spectral configurations among different sensors, fine-tuning all layers without partial freezing was found to accelerate convergence, as evidenced by a faster decrease in the loss. During the fine-tuning process, most training settings were kept consistent with full training; however, the number of epochs was reduced to 30, and the learning rate was decreased by a factor of 10 if the validation loss did not decrease for 3 consecutive epochs.

### 3.3 The overall RIFTES framework for LST retrieval

The RTM-free atmospheric correction model based on neural networks and the non-iterative TES solver constitutes the RIFTES framework, with the corresponding workflow shown in Fig. 4. Ground-leaving and atmospheric downward radiances were first estimated from TOA radiances, VZA, and optional PWV using DRNNMM. Subsequently, the non-iterative TES algorithm was applied to simultaneously retrieve LST and LSE in a direct closed-form manner. Table 2 summarizes the requirements of input parameters and iterative procedures for different algorithms, including SW, TES, a representative hybrid method termed the improved SW-TES (ISW-TES) method (Wang et al., 2025), and RIFTES. The new framework enables clear-sky LST retrieval without relying on prior information (e.g., atmospheric profiles and LSE) and intensive computations (e.g., from the atmospheric RTM and multiple iterations in the NEM module), which is expected to improve both the retrieval accuracy and computational efficiency.

**Table 2.** Comparison of requirements of iterative procedures and input parameters for different clear-sky LST retrieval algorithms.

| Method | Iterative procedures | Input requirements | | |
|---|---|---|---|---|
| | | Atmospheric profiles | PWV | LSE |
| SW | No | No | Yes | Yes |
| TES | Yes | Yes | No | No |
| ISW-TES | Yes | No | Yes | No |
| RIFTES | No | No | Optional | No |

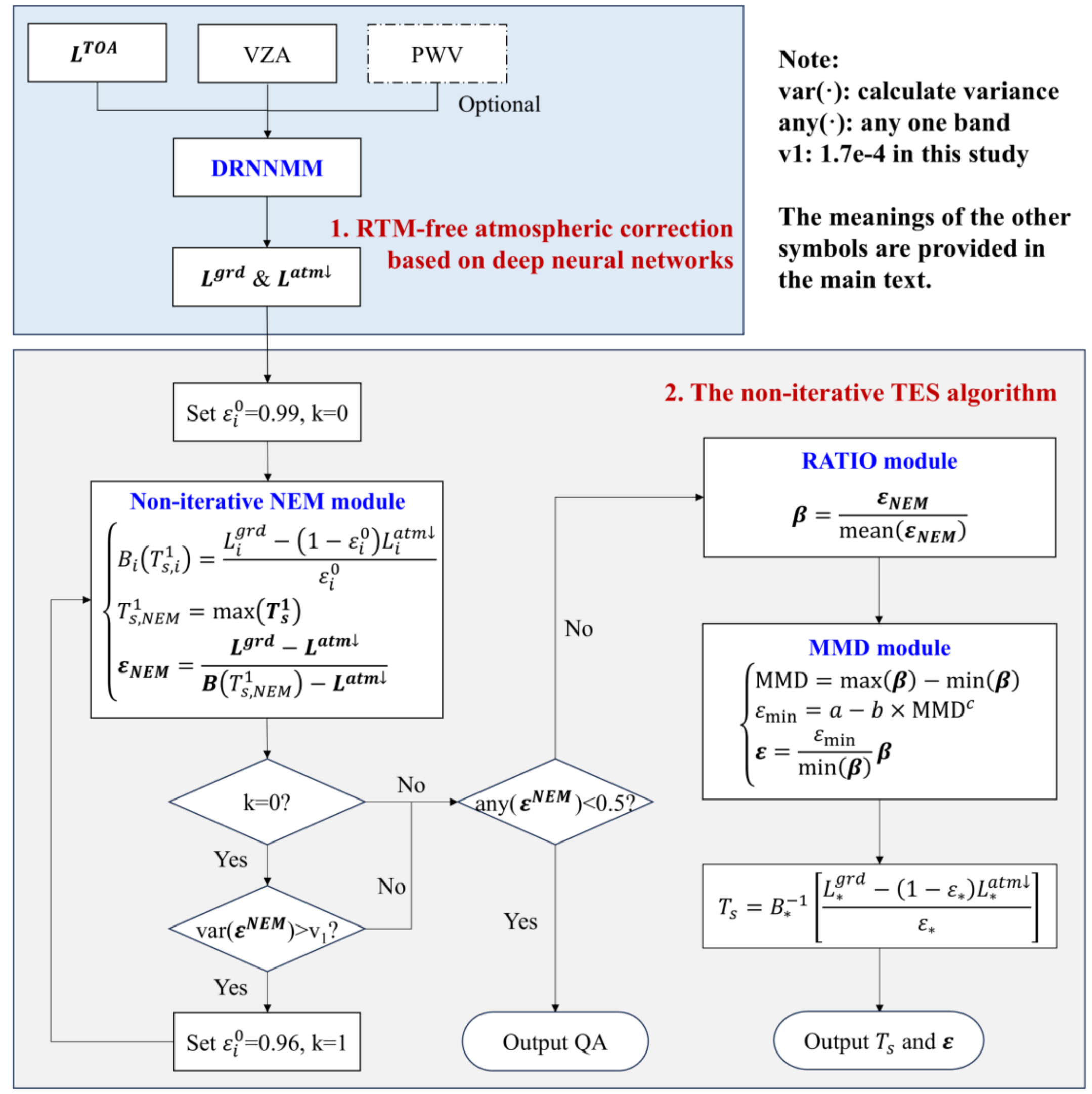

**Fig. 4.** Overall workflow of the RIFTES framework for clear-sky LST retrieval.

### 3.4 Evaluation strategies

Comprehensive simulation analysis and T-based validation were conducted in this study to validate the performance of various clear-sky LST retrieval methods, including SW, TES, RIFTES. These two evaluation strategies complement each other to provide more robust and credible validation results, as simulation datasets can provide sufficient samples with global representativeness, while in-situ measurements reflect the accuracy of methods when applied to real satellite observations.

#### 3.4.1 Simulation analysis

The simulation analysis was conducted based on the simulation dataset compiled for ECOSTRESS. The effectiveness of the non-iterative TES algorithm was first evaluated by comparing the initial LST and LSE estimates obtained from the conventional NEM module with those from the non-iterative NEM module, as well as the final LST estimates obtained from the conventional TES algorithm with those from the non-iterative TES algorithm.

Subsequently, the accuracies of the four algorithms were compared, with the SW method based on the formulation proposed by Jimenez-Munoz and Sobrino (2008). The coefficients of the SW equations in both the SW and ISW-TES algorithms, as well as the model parameters of DRNNMM, were first determined using the simulated training set, after which the accuracies of the four algorithms were evaluated on the simulated test set. The input parameters of all algorithms were initially assumed to be error-free to reflect the theoretical accuracy. Then, to evaluate models' performance under real-world scenarios, the input parameters were perturbed with random errors according to their typical uncertainty levels. Specifically, Gaussian errors with a standard deviation (STD) of 0.015 W $m^{-2}$ $sr^{-1}$ $\mu m^{-1}$ were introduced to TOA radiances involved in all algorithms (Zhang et al., 2025a). PWV inputs required by the SW, ISW-TES, and RIFTES algorithms were perturbed with Gaussian errors with an STD of 4% (Zhang et al., 2025a). LSE inputs required by the SW algorithm were added with Gaussian errors with an STD of 0.01 (Cheng et al., 2025). Atmospheric profiles required by the TES algorithm were also perturbed, with the humidity profiles adjusted using random scaling factors uniformly distributed between 0.8 and 1.2, while the temperature profiles were perturbed with Gaussian errors having an STD of 2 K (Hulley et al., 2012; Liu et al., 2022; Zhang et al., 2025b).

### 3.4.2 T-based validation

The T-based validation method was conducted based on spatiotemporally matched satellite retrievals and in-situ measurements to assess the practical performance of various algorithms, with the algorithms calibrated using the corresponding simulated training datasets when necessary.

For ECOSTRESS, LST estimates from NASA's official ECO_L2_LSTE v002 product were directly used as the estimation results of the conventional TES algorithm, while the other three methods were independently reproduced. For simplicity, LSEs for Bands 4 and 5 required by the SW method were directly derived from the ECO_L2_LSTE v002 product. Observations labeled as "confident clear" or "probably clear" in the ECO_L2_CLOUD v002 product and identified as having best or good geolocation accuracy in the ECO_L1B_GEO v002 product were retained. Considering the accuracy of cloud detection results may be limited as ECOSTRESS relies solely on TIR bands for cloud masking, an additional criterion was introduced to further minimize the influence of cloud contamination, which requires LSE estimates of ECOSTRESS Band 5 from the ECO_L2_LSTE v002 product over gray bodies be greater than 0.95 (Hulley et al., 2022; Zhang et al., 2025c). For GOES-16, LST estimates from NOAA's official LST product were directly used as the estimation results of the SW algorithm, while the other three methods were independently reproduced. Only observations identified as clear-sky in NOAA's official LST product were retained.

After selecting clear-sky observations for each sensor, the "3σ-Hampel identifier" method (Göttsche et al., 2013) was applied to each algorithm to remove possible outliers caused by undetected clouds or instrumental disturbances. The intersection of the retained samples from all algorithms for each sensor was then used to calculate the accuracy metrics, including the root mean squared error (RMSE) and bias.

# 4. Results

## 4.1 Simulation analysis

### 4.1.1 Comparison of the conventional and non-iterative TES algorithms

A simulation analysis was first conducted to demonstrate the effectiveness of the new non-iterative TES algorithm. Specifically, both the conventional and non-iterative TES algorithms were applied to the simulation dataset for ECOSTRESS, estimating LST from error-free $\boldsymbol{L^{grd}}$ and $\boldsymbol{L^{atm\downarrow}}$. As $\boldsymbol{L^{grd}}$ and $\boldsymbol{L^{atm\downarrow}}$ are not changed with VZA, thus only a subset of the complete simulation dataset containing 1,690,725 samples (67,629 groups of atmospheric parameters × 5 LSTs × 5 LSEs) were used here.

For each sample, the maximum LST obtained during each iteration of the NEM module was compared. The results showed that, for all samples in the simulation dataset, the maximum LST was determined in the first iteration and remained unchanged throughout the subsequent iterations. For better visualization, four samples were random selected to illustrate the iterative process of the conventional NEM module, with the initial LST and LSE estimates presented in Figs. 5 and 6, respectively. For visualization purposes, the maximum number of iterations was fixed at 30, and no convergence test was applied. As shown in Fig. 5, the NEM LST remains unchanged throughout the iterative procedure, while the LSTs of the other bands increase progressively and gradually converge toward the NEM LST. For the same four samples, Fig. 6 further compares the LSE estimates obtained from the conventional and new non-iterative NEM modules. Similarly, LSE of the band corresponding to the maximum LST in the NEM module remains unchanged, whereas the LSEs of the other bands increase iteratively and gradually converge toward those derived from the non-iterative NEM module. Therefore, these experimental results are consistent with the physical derivation results presented in Eqs. (9), (13), and (15).

Fig. 7 further compares the final LST estimates from the conventional and new non-iterative TES algorithms. As the initial LSE values derived from the NEM and non-iterative NEM modules are nearly identical, the differences in the final LST estimates after the RATIO and MMD modules are also negligible (i.e., RMSE=0.00 K, bias=0.00 K). Therefore, based on both the experimental results in this section and the physical derivations in Section 3.1, the non-iterative TES algorithm produces results identical to those of the conventional TES algorithm while eliminating the multiple iterations required by the traditional NEM module, thereby serving as an effective alternative to the conventional TES algorithm.

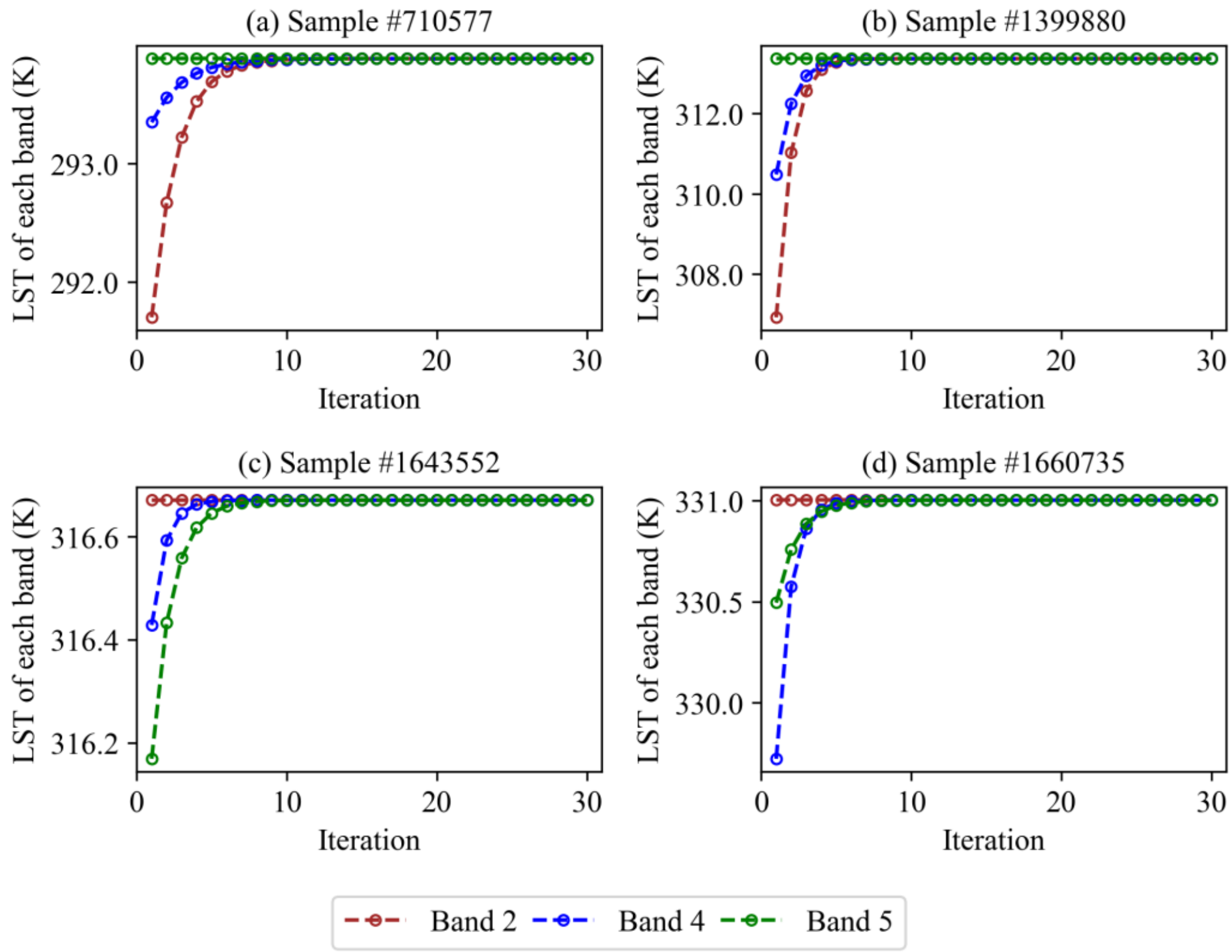

**Fig. 5.** LST estimates for each band during the NEM iteration processes for four random samples in the simulation dataset compiled for ECOSTRESS.

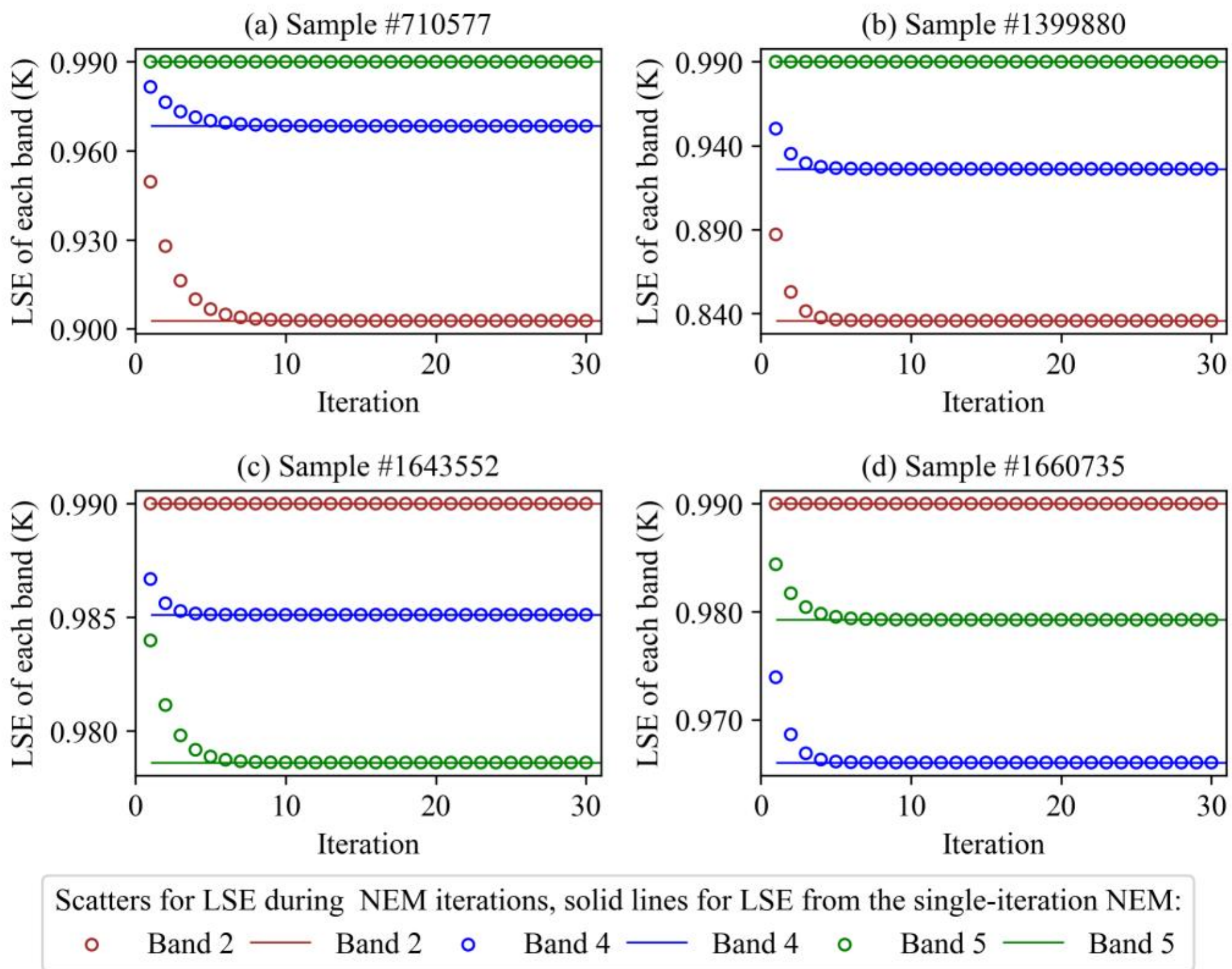

**Fig. 6.** LSE estimates from the conventional and non-iterative NEM modules for four random samples in the simulation dataset compiled for ECOSTRESS.

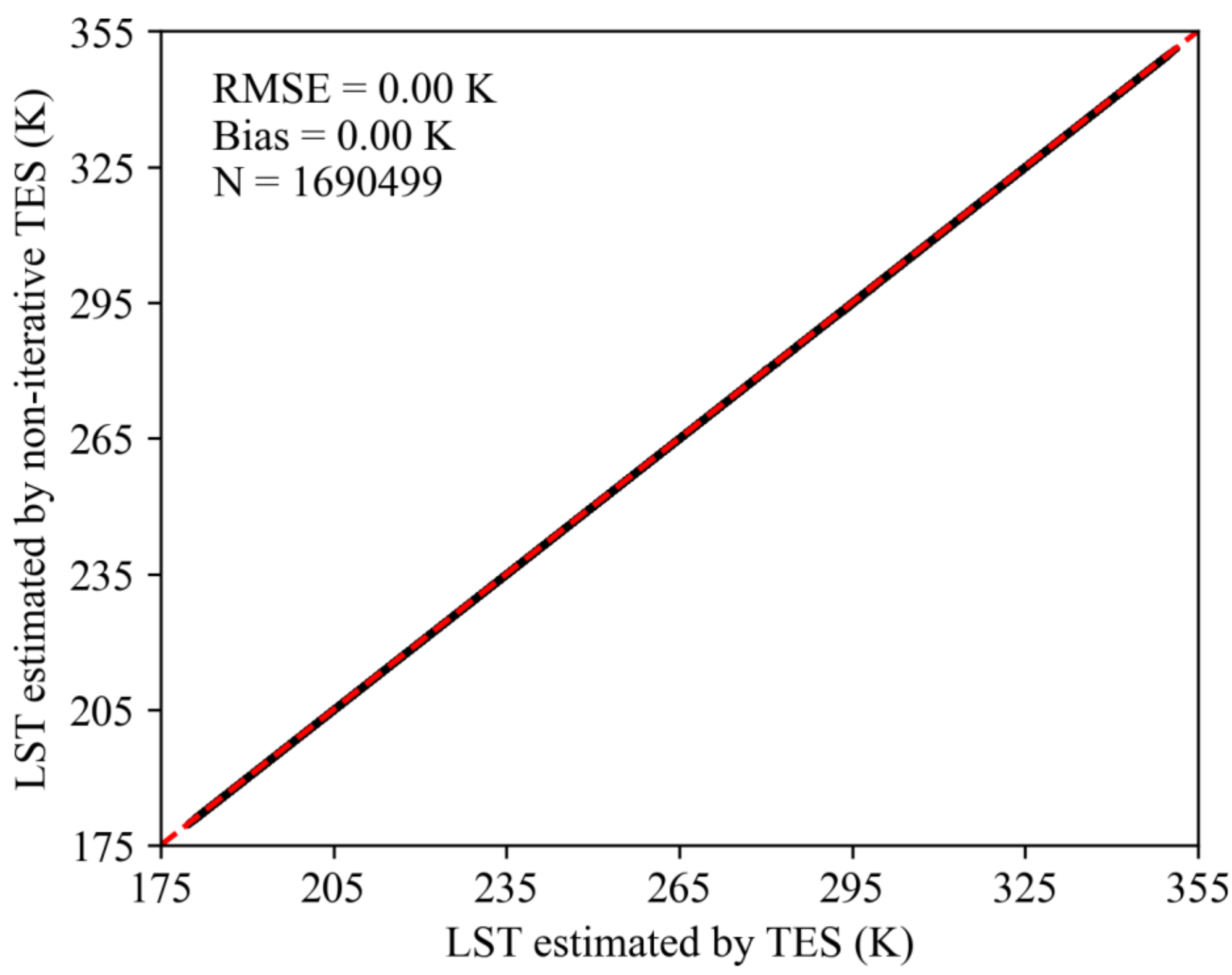

**Fig. 7.** Comparison between LST estimates from the conventional and non-iterative TES algorithms on the simulated dataset compiled for ECOSTRESS.

### 4.1.2 Performance of different algorithms

Fig. 8 provides the distribution of LST estimation errors for different algorithms (i.e., SW, TES, ISW-TES, and RIFTES) on the simulated test set compiled for ECOSTRESS. The biases of all algorithms are close to zero except the TES algorithm with input uncertainties (bias=0.57 K). For the SW algorithm, its RMSE increases from 0.99 K under ideal conditions to 1.90 K when uncertainties in the input parameters are introduced. This accuracy degradation is primarily attributable to errors in LSE (Cheng et al., 2025). For the TES algorithm, as the atmospheric RTM was assumed to be accurate during simulation analysis, the performance of TES without uncertainties (RMSE=0.58 K) is best among all algorithms. However, considering the uncertainties arising from spatial and temporal mismatches between remote sensing observations and atmospheric profiles, as well as those associated with the atmospheric RTM, the ideal performance of TES may be somewhat optimistic. When adding uncertainties to input parameters, the RMSE of TES increases rapidly to 3.51 K, with a pronounced overestimation generally observed in high-temperature regions (Zhang et al., 2025b). Limited by the modeling capability of conventional SW methods, the ISW-TES algorithm achieves an ideal RMSE of 1.84 K, which slightly increases to 1.87 K after uncertainties are introduced into the input parameters. In comparison, the RIFTES algorithm (incorporates PWV for atmospheric correction) achieves an ideal RMSE of 1.01 K, with the RMSE increasing only marginally to 1.06 K under perturbed conditions. Even without PWV as an input and relying solely on remote sensing observations, the proposed algorithm remains stable with minor accuracy degradation (i.e., RMSE=1.38 K/1.42 K without/with input uncertainties, respectively). A possible explanation is that, among the input parameters, the observation difference between adjacent TIR bands (i.e., $dL_{2,3}^{TOA}$) is closely related to the atmospheric absorption effect, and cosVZA can also provide information related to atmospheric effects (Dalu, 1986). Therefore, DRNNMM can extract the atmospheric information embedded in these input parameters, thereby reducing the reliance on PWV to some extent.

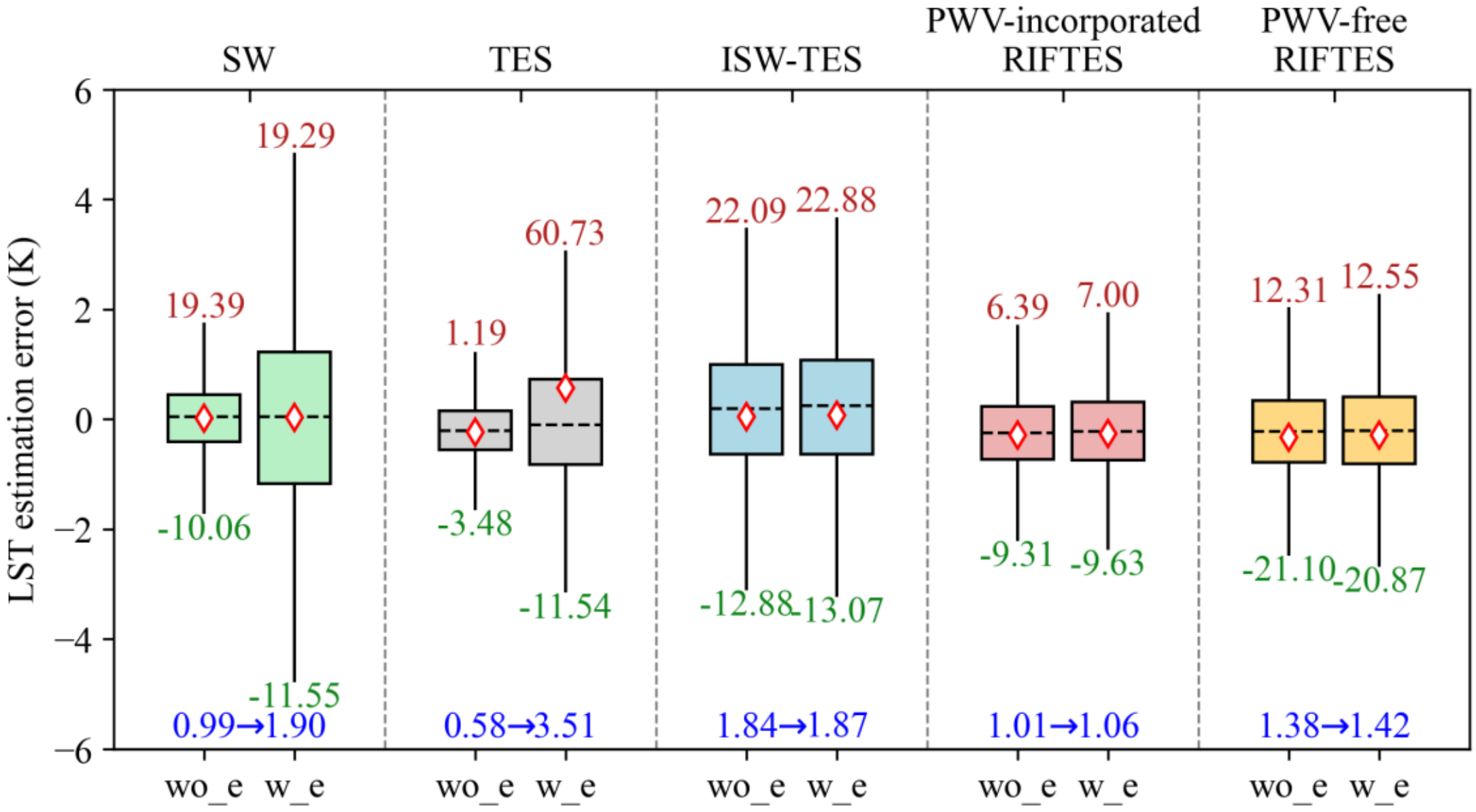

**Fig. 8.** Validation results of four LST retrieval algorithms over the simulated test set compiled for ECOSTRESS. ‘wo_e’ and ‘w_e’ mean ‘without errors in input parameters’ and ‘with error in input parameters’, respectively. For better visualization, outliers of the boxplots are omitted. The red text above the upper whisker represents the maximum error value, while the green text below the lower whisker represents the minimum error value. The blue text represents RMSE. Within each box, the black dashed line and the red diamond represent the median value and mean value, respectively.

Fig. 9 further compares the RMSE values of the PWV-incorporated and PWV-free RIFTES algorithms across different LST and PWV bins when uncertainties are included in the input parameters. The accuracy differences between these algorithms are primarily observed under extremely hot and humid conditions with relatively small sample sizes, whereas the differences remain relatively minor under other conditions.

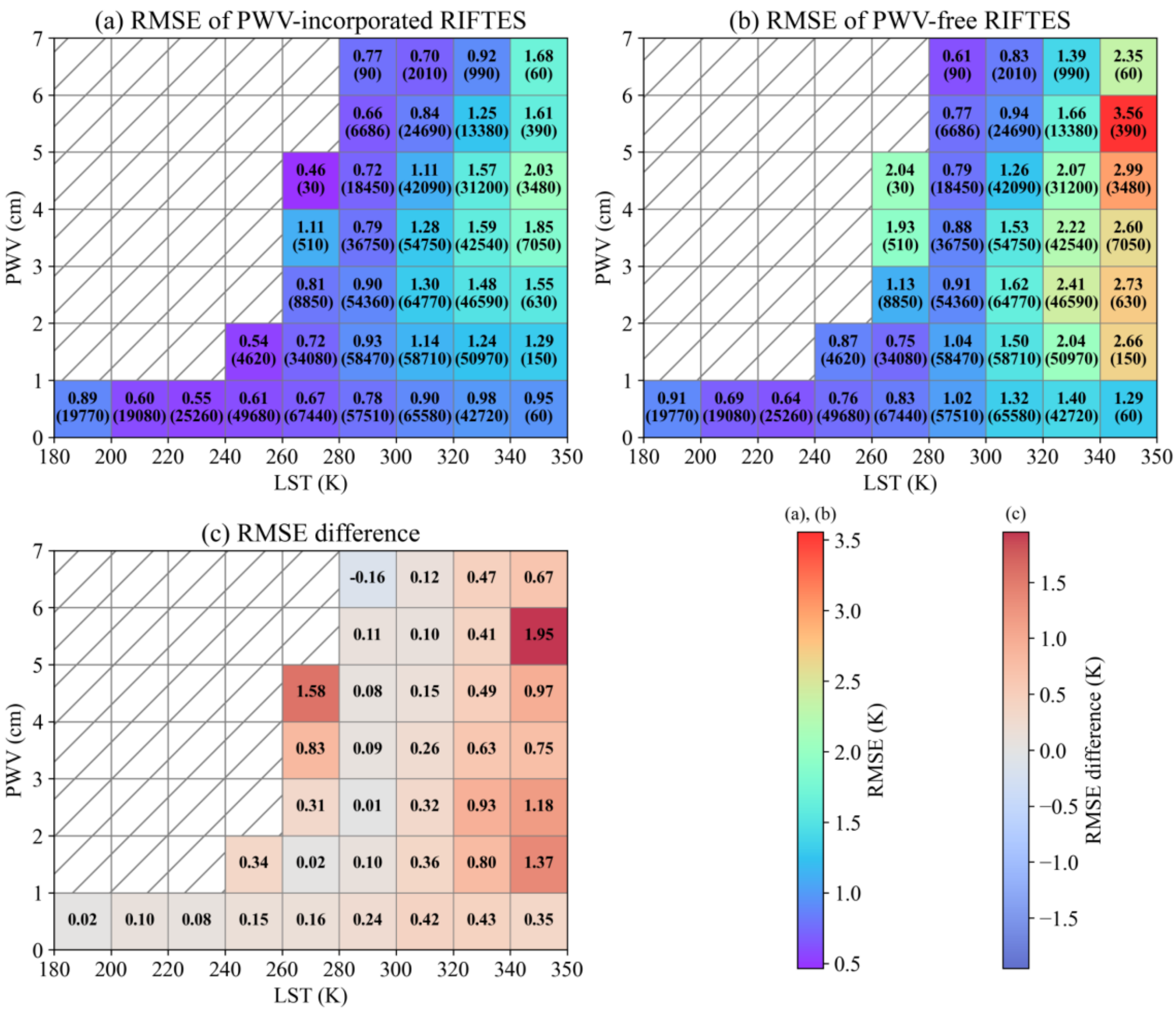

**Fig. 9.** Comparison of RMSE values of the PWV-incorporated and PWV-free RIFTES algorithms across different LST and PWV bins. Panels (a) and (b) present RMSE values of the PWV-incorporated and PWV-free RIFTES algorithms. Numbers in parentheses indicate the number of samples. Panel (c) shows the RMSE differences (PWV-free RIFTES minus PWV-incorporated RIFTES).

### 4.2 T-based validation

As the SW, TES, and ISW-TES algorithms generally require prior atmospheric information such as PWV or atmospheric profiles, the RIFTES algorithm in this section also incorporates PWV as an input to ensure a fair comparison.

Based on the in-situ measurements from the five radiometer sites, Fig. 10 presents the T-based validation results of different algorithms when applied to the ECOSTRESS sensor. The irregular orbit of the ECOSTRESS sensor leads to limited overpasses over the Iberian Peninsula, yielding fewer validation samples at the CDG and FDU sites. Overall, LST estimates from all methods show good agreement with in-situ measurements. The RIFTES and TES algorithms achieve RMSE values below 2 K at four sites, whereas the other two methods meet this criterion at three sites. The biases of RIFTES at all five sites range from -0.07 K to 0.25 K, demonstrating its stable retrieval performance over different regions. The ISW-TES algorithm appears to underestimate LST at the barren GBBp site (bias=-1.35 K), which may be attributable to limitations of the SW algorithm (Zhang et al., 2025a; Zheng et al., 2022). The overall RMSEs of SW, TES, ISW-TES, and RIFTES are 1.98 K, 1.64 K, 1.91 K, and 1.51 K, respectively. Compared with the other three methods, the proposed RIFTES algorithm reduces the overall retrieval uncertainty by 24%, 8%, and 21%, respectively.

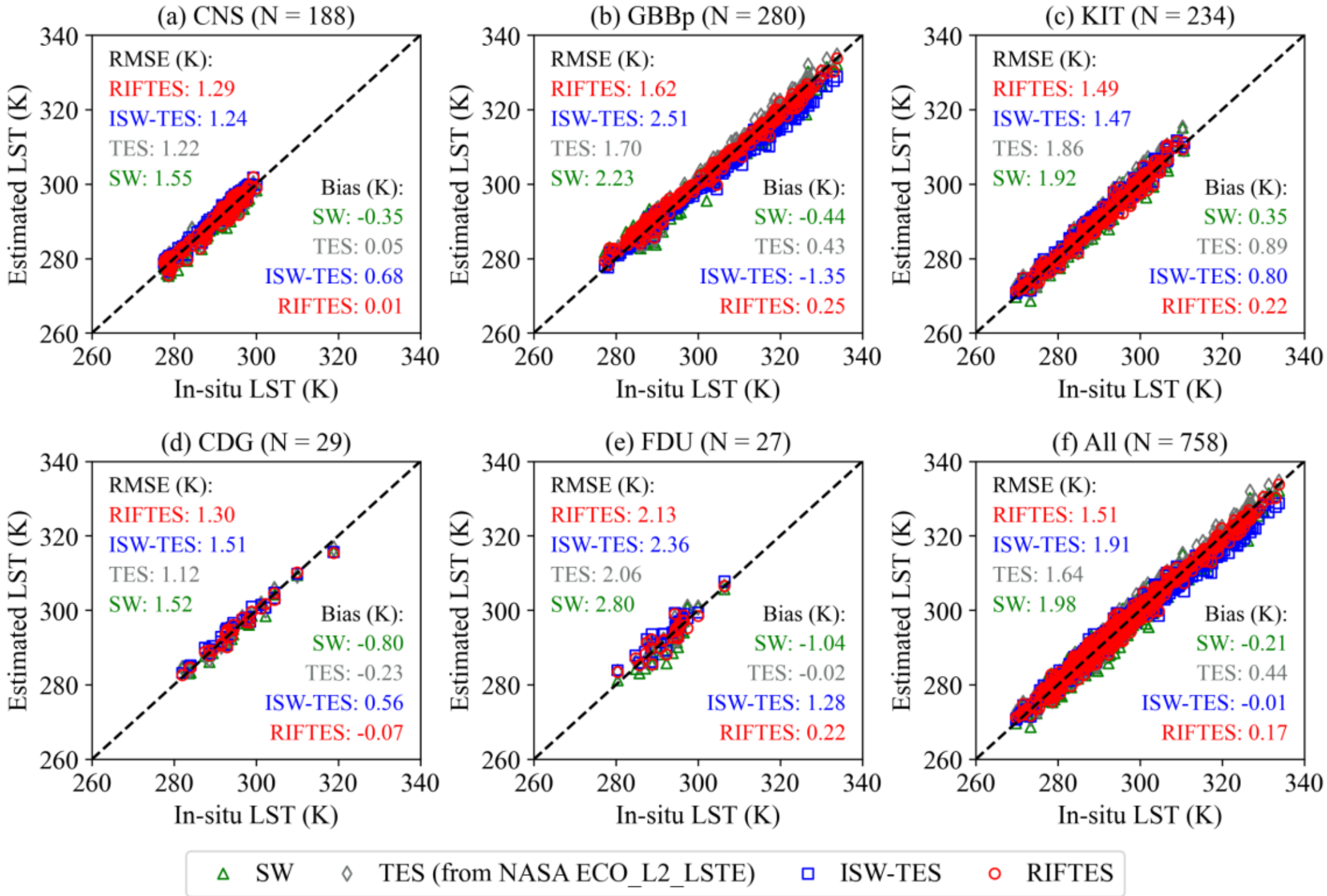

**Fig. 10.** T-based validation results of different algorithms on radiometer sites for the ECOSTRESS sensor.

The RIFTES algorithm was then applied on GOES-16 ABI for estimating hourly, 2 km clear-sky LST. Fig. 11 presents the corresponding T-based validation results over the seven SURFRAD sites. Benefiting from the high temporal sampling frequency of geostationary satellites, each site contains several thousand validation samples, with the total number exceeding 23,000. The RIFTES algorithm achieves RMSE values below 2 K at four sites, whereas the SW, TES, and ISW-TES algorithms achieve this level of accuracy at three, zero, and three sites, respectively. At the barren DRA site, the SW and ISW-TES algorithms exhibit relatively noticeable underestimation biases, and the SW algorithm yields an RMSE of only 4.17 K, which may be caused by inaccurate emissivity priors. The biases of all methods at the TBL site exceed -1.4 K, which may be attributed to the fact that this site is located in mountain areas with pronounced spatial heterogeneity (Jiménez et al., 2017; Meng et al., 2022; Sekertekin, 2019). When applied to GOES-16 ABI, the RIFTES algorithm again achieves the lowest overall RMSE (1.97 K) among the four methods, reducing 28%, 32%, and 14% retrieval uncertainty compared to the SW (2.75 K), TES (2.88 K), and ISW-TES (2.28 K) algorithms, respectively. It should be noted that the reported accuracies of the SW and ISW-TES algorithms are very close to those reported by Wang et al. (2025), although the ISW-TES algorithm in this study uses a different channel configuration. The validation accuracies for GOES-16 ABI are generally lower than those for ECOSTRESS, which may be attributed to several factors: (1) Radiometer sites are of higher quality and exhibit better consistency with satellite-derived LST in general; (2) The pixel size of ECOSTRESS (about 70 m) is more consistent with the effective representativeness scale of in-situ measurements, while the spatial resolution of ABI (2 km at nadir) may be too coarse; (3) ABI observations over SURFRAD sites are associated with larger VZA, which typically introduce greater atmospheric correction uncertainties.

To further evaluate the robustness of these four methods under different conditions, Fig. 12 compares the corresponding RMSE values under daytime and nighttime conditions, as well as under extremely humid and hot conditions (PWV and LST exceeding the 90th percentile at each site). The RIFTES algorithm exhibits greater robustness across different sensors and conditions, with maximum RMSE values below 1.7 K for ECOSTRESS and below 2.5 K for GOES-16 ABI, respectively. In contrast, other methods may experience significant accuracy degradation under specific conditions. For example, when applied to ECOSTRESS, the RMSE of ISW-TES reaches 2.95 K under high LST conditions; when applied to GOES-16 ABI, the RMSE values of SW/TES are 2.97/3.04 K during daytime, 3.55/3.94 K under high PWV conditions, and 4.47/4.10 K under high LST conditions.

Therefore, by integrating the powerful learning capability of DL models with the multi-source physical foundations from the atmospheric RTM and conventional clear-sky LST retrieval algorithms, the RIFTES algorithm generally achieves higher retrieval accuracy and remain robust under different conditions. It also effectively eliminates the reliance on both LSE information and atmospheric profiles, thereby reducing the risk of error propagation.

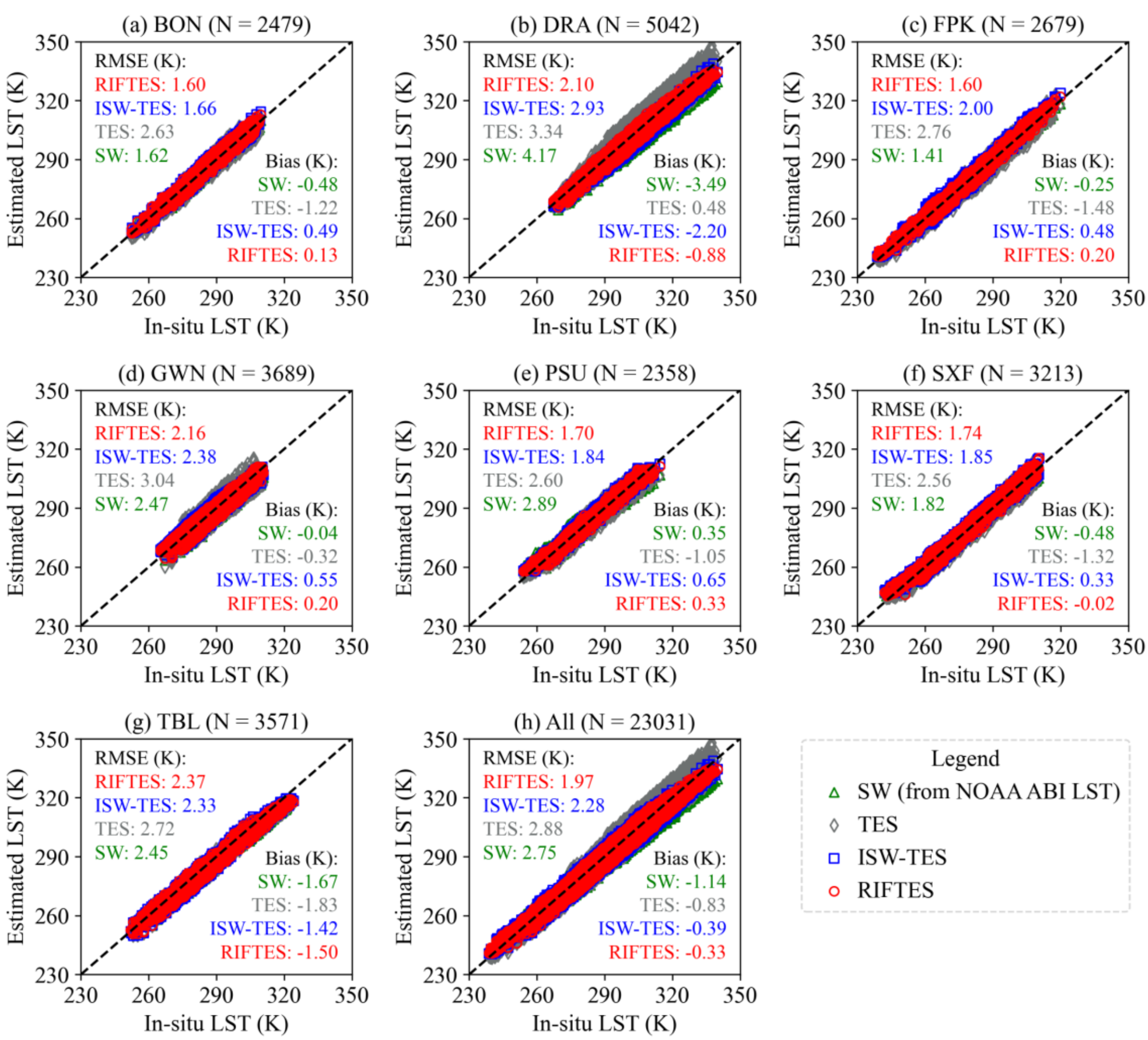

**Fig. 11.** T-based validation results of different algorithms on pyrgeometer sites for the GOES-16 ABI sensor.

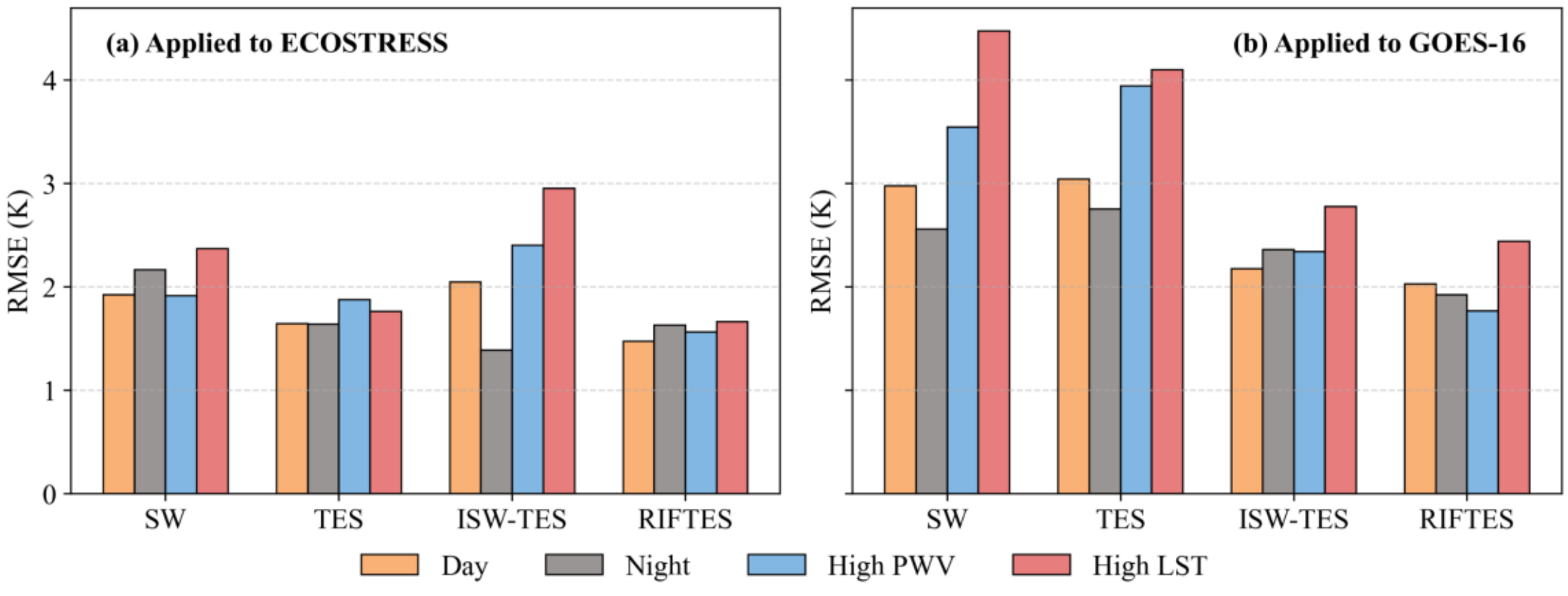

**Fig. 12.** Performance of four algorithms under different sensors and conditions.

### 4.3 Comparison of computational efficiency

In addition to retrieval accuracy, computational efficiency is also a key factor that should be considered in practical applications of the LST retrieval algorithm. In this section, the computational efficiency of TES and TES-based hybrid methods (i.e., ISW-TES and RIFTES) was evaluated using multiple ECOSTRESS scenes, with the average computational time recorded for comparison. The experiment was conducted on the same device equipped with a 12th Gen Intel Core i9-12900H CPU (2.50 GHz), an NVIDIA GeForce RTX 3070 Ti Laptop GPU (8 GB), and 32 GB RAM.

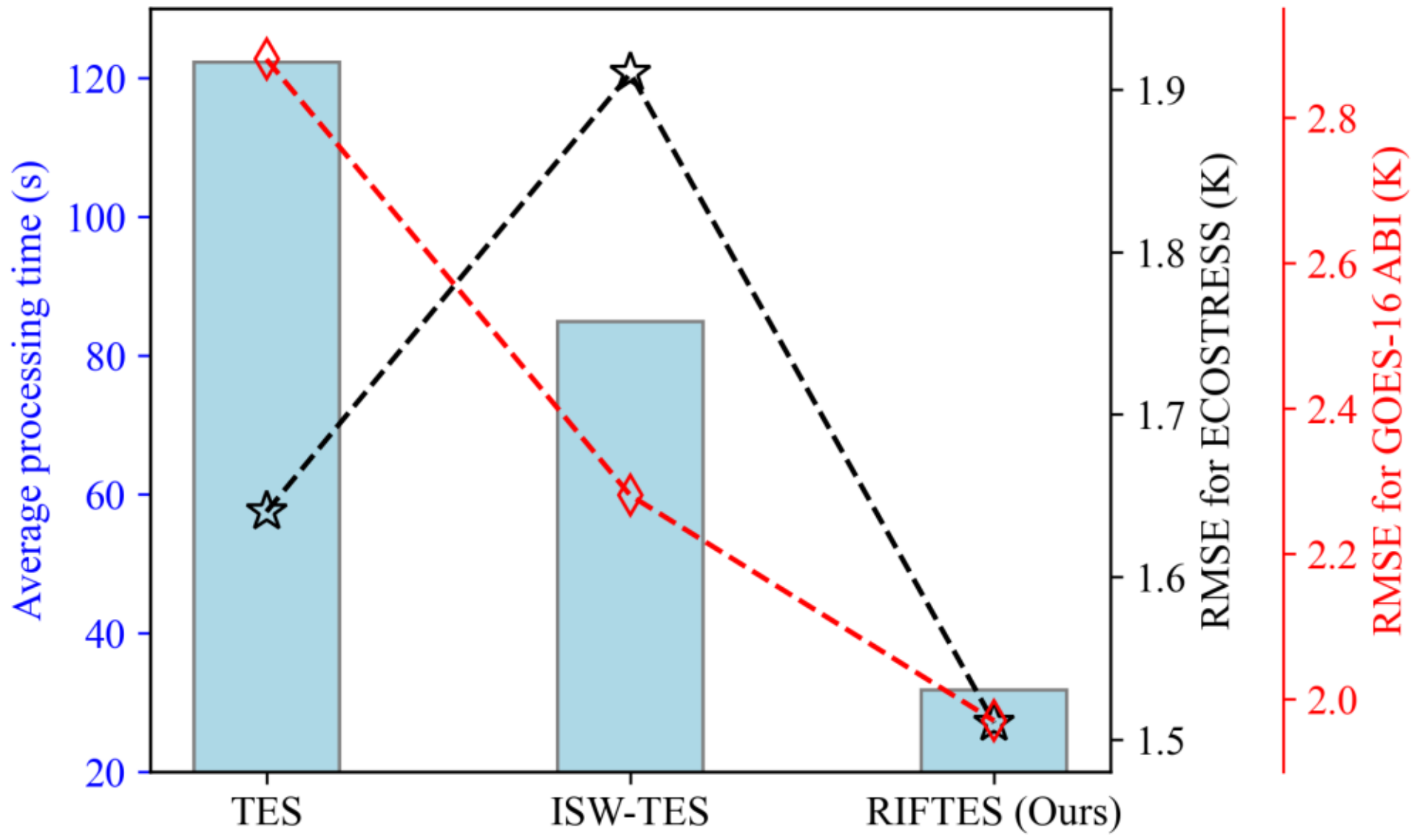

**Fig. 13.** Comparison of computational efficiency and retrieval accuracy among different methods. Accuracy metrics are derived from the T-based validation results in Section 4.2.

Fig. 13 presents the comparison of processing time and retrieval accuracy among different methods, with the accuracy metrics taken from Section 4.2. According to the experimental results, the average computational speed of the TES, ISW-TES, and RIFTES algorithms were 122.29, 84.82, and 31.78 seconds per scene, respectively. The TES algorithm required the longest computation time because it involves both intensive radiative transfer calculations for atmospheric correction and multiple iterations in the NEM module. The ISW-TES algorithm simplifies the atmospheric correction step, and its iterative convergence in the subsequent NEM module was found to be generally faster than that of the traditional TES algorithm, which may be attributed to differences in the atmospheric correction schemes. However, the iterative procedure involved in the NEM module still demand substantial computation, resulting in relatively high computational cost of ISW-TES. In comparison, by simplifying both the atmospheric correction step and the NEM module, the RIFTES algorithm effectively reduces the computational time by 74.0% and 62.5% compared to the conventional TES and ISW-TES algorithms, respectively. Although the absolute runtime is expected to decrease on high-performance computing platforms, the relative computational efficiency among the three methods is generally expected to remain similar.

Therefore, the RIFTES algorithm not only achieves better retrieval accuracy and robust performance, but also effectively reduces the dependence on atmospheric parameters and improves the computational efficiency. Such a dual-benefit characteristic distinguishes this new algorithm from other approaches developed to refine the TES algorithm. For example, the WVS algorithm is a representative method for improving the accuracy of atmospheric correction, but at the cost of largely increased computational complexity. Consequently, the computational efficiency gap between the RIFTES and WVS algorithms is expected to be even more pronounced. For another example, most hybrid methods that combine the conventional SW and TES algorithms may not provide substantial improvements in either retrieval accuracy or computational efficiency. In this context, the new RIFTES framework may exhibit promising potential for practical operational deployment.

# 5. Discussion

## 5.1 Impact of the number of residual blocks in DRNNMM

The performance of neural networks generally varies with model capacity. To further assess its influence, the simulated training set for ECOSTRESS was used to train DRNNMM with the number of ResBlocks varying from 1 to 4. Subsequently, the corresponding loss values of DRNNMM with different model depths for estimating $\boldsymbol{L^{grd}}$ and $\boldsymbol{L^{atm\downarrow}}$ were calculated on the test set using Eq. (21). Table 3 compares the corresponding loss values of PWV-incorporated and PWV-free DRNNMM with different model depths. The loss values of the PWV-free DRNNMM are significantly larger than those of the PWV-incorporated model, mainly due to the larger errors in the predicted $\boldsymbol{L^{atm\downarrow}}$ when PWV is not included as an input (Zhang et al., 2025a). Overall, the variations in loss values across different model depths are relatively small, indicating strong learning capability and good stability of the model. As the network depth increases, model performance tends to saturate, with the lowest loss achieved when three ResBlocks are used. Considering that DRNNMM is already a highly lightweight network, increasing or decreasing the number of ResBlocks has only a limited impact on model complexity. Therefore, three ResBlocks with the best performance are ultimately adopted.

**Table 3.** Loss values of PWV-incorporated and PWV-free DRNNMM with different numbers of ResBlocks on the simulated test set compiled for ECOSTRESS.

| Number of ResBlocks | Loss values | |
|---|---|---|
| | Include PWV | Exclude PWV |
| 1 | 0.0408 | 0.4297 |
| 2 | 0.0389 | 0.4160 |
| 3 | 0.0385 | 0.4133 |
| 4 | 0.0386 | 0.4134 |

## 5.2 Comparison with an explicitly physics-embedded DL method

The DRNNMM model directly estimates $\boldsymbol{L^{grd}}$ and $\boldsymbol{L^{atm\downarrow}}$ in an end-to-end way, in which the physical mechanisms from atmospheric RTMs and the SW method were incorporated into the DL model through an implicit way. In recent years, some studies attempted to embed the SW equation into the loss function to explicitly integrate physical mechanisms into DL models for LST retrieval (Cheng et al., 2025). Following the similar idea, this section further proposes a SW-embedded DL (SWEDL) model to estimate grounding-leaving brightness temperature, which is then cascaded with the non-iterative TES algorithm to form the SWEDL-TES algorithm for LST retrieval.

According to the physical deviation results of Zheng et al. (2019) as well as the best band combination scheme for ECOSTRESS reported by Zhang et al. (2025a), the SW method for estimating grounding-leaving brightness temperature was written as follows:

$$\begin{cases} \hat{T}_1^{grd} = A_0 + A_1 T_1^{TOA} + A_2(T_1^{TOA} - T_3^{TOA}) + A_3(T_1^{TOA} - T_3^{TOA})^2 \\ \hat{T}_2^{grd} = A_4 + A_5 T_2^{TOA} + A_6(T_2^{TOA} - T_3^{TOA}) + A_7(T_2^{TOA} - T_3^{TOA})^2 \\ \hat{T}_3^{grd} = A_8 + A_9 T_3^{TOA} + A_{10}(T_3^{TOA} - T_2^{TOA}) + A_{11}(T_3^{TOA} - T_2^{TOA})^2 \end{cases} \tag{22}$$

where $\hat{T}_1^{grd}$, $\hat{T}_2^{grd}$, and $\hat{T}_3^{grd}$ represents the estimated grounding-leaving brightness temperature for ECOSTRESS Bands 2, 4, and 5, $T_1^{TOA}$, $T_2^{TOA}$, and $T_3^{TOA}$ denotes the corresponding TOA brightness temperature, and $A_i$ represents the coefficients.

Different from the traditional SW method that assumes constant coefficients, the SWEDL model estimates $A_i$ through neural networks using input parameters similar to those of DRNNM:

$$A_i = \mathcal{g}(T_1^{TOA}, T_2^{TOA}, T_3^{TOA}, \cos\theta_v, \mathrm{PWV}) \tag{23}$$

where $\mathcal{g}$ denotes the SWEDL model.

The predicted coefficients were then input into Eq. (22) to estimate $\hat{\boldsymbol{T}}^{\boldsymbol{grd}}$, which was subsequently compared with the ground-truth labels (i.e., $\boldsymbol{T}_{\boldsymbol{true}}^{\boldsymbol{grd}}$) for loss computation and model optimization:

$$\mathcal{L}' = \left\| \boldsymbol{T}_{\boldsymbol{true}}^{\boldsymbol{grd}} - \hat{\boldsymbol{T}}^{\boldsymbol{grd}} \right\|^2 \tag{24}$$

The SWEDL model was trained using the same configurations for the DRNNMM as mentioned in Section 3.2.2 (the strategy of randomly dropping PWV was not adopted). The SWEDL model can only estimate ground-leaving brightness temperature, which was converted into $\boldsymbol{L}^{\boldsymbol{grd}}$ using the Planck function. To drive the non-iterative TES algorithm for LST retrieval, $\boldsymbol{L}^{\boldsymbol{atm}\downarrow}$ was further obtained from the DRNNMM model for simplicity, as its generally has a limited impact on LST retrieval.

Fig. 14 compares the retrieval accuracy of the SWEDL-TES and RIFTES algorithms on both the simulated test set and in-situ measurements. As shown in Fig. 14(a), the discrepancies between the LST estimates from the two methods are generally minor, resulting in nearly identical RMSE values (1.06 K versus 1.05 K) and bias values (both -0.26 K) on the simulated test set. Fig. 14(b) shows the T-based validation results, in which the LST estimates from the ECO_L2_LSTE product was also included for comparison. As the “3σ-Hampel identifier” method was applied to different algorithms, the number of samples as well as reported accuracy metrics of TES and RIFTES slightly differ from that in Fig. 10. The overall RMSE values of the RIFTES and SWEDL-TES algorithms are 1.57 K and 1.62 K, respectively. Therefore, the two algorithms still achieve comparable performance in practical applications, with RIFTES yielding a slightly lower RMSE.

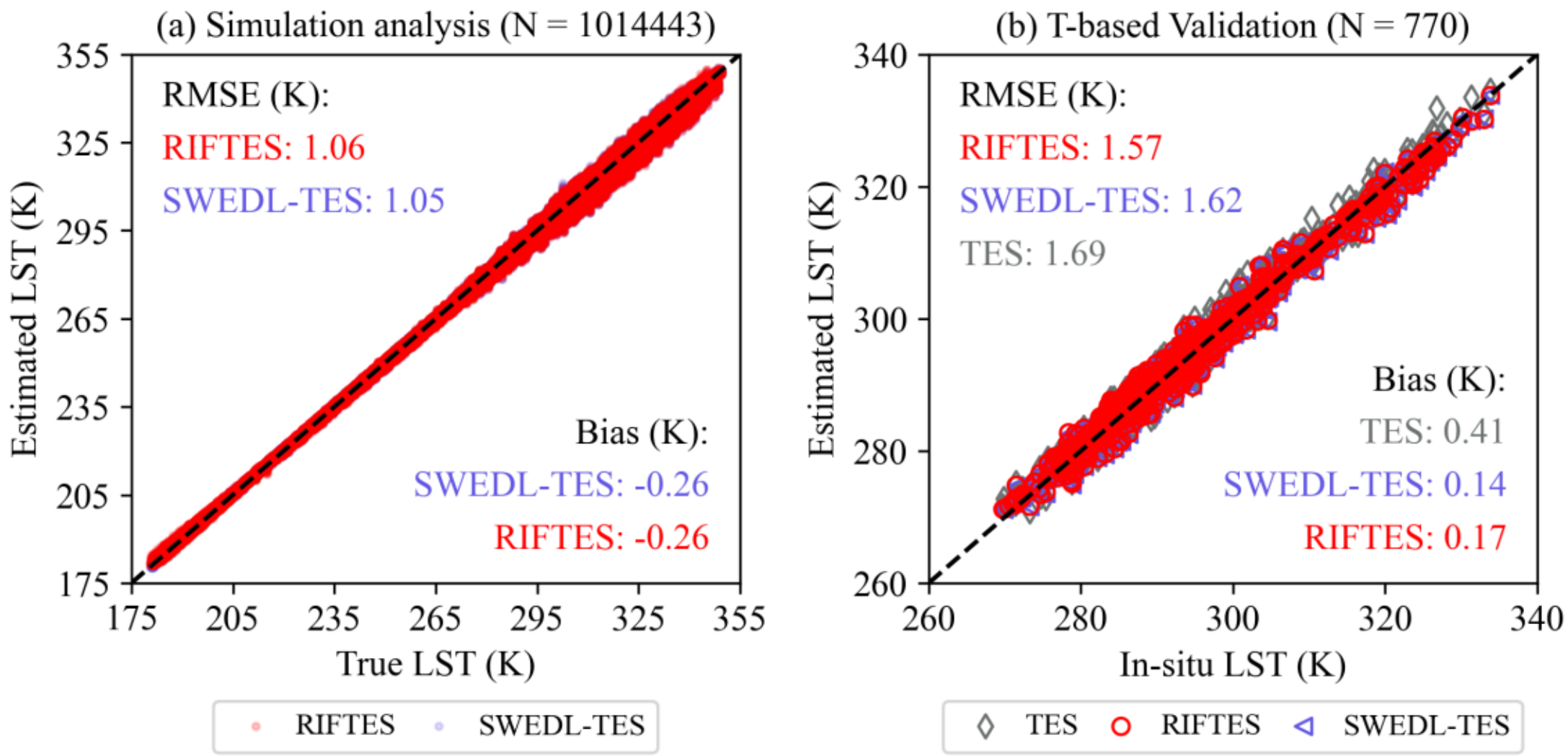

**Fig. 14.** Performance of the RIFTES and SWEDL-TES algorithms over (a) the simulation test set and (b) in-situ measurements.

According to the above analyses, the SWEDL-TES method performs similarly to the RIFTES algorithm, which may be attributed to the following two reasons. First, the simulation dataset used for model training was constructed based on atmospheric radiative transfer mechanisms and covers a wide range of atmospheric and land surface conditions. Therefore, trained using this physically derived and sufficiently representative dataset, the neural network is able to effectively capture the underlying physical relationships in an implicit manner, achieving accurate atmospheric correction results without additional constraints imposed by the loss function. In fact, the SWEDL model may offer advantages in small-sample scenarios (Cheng et al., 2025; Guan et al., 2026). However, considering that constructing a comprehensive simulation dataset for model training is not particularly challenging, this advantage may be less critical in practice. Second, the SW method for estimating ground-leaving brightness temperature also involves assumptions and approximations during its physical derivation, which may be difficult to fully represent the exact physical principles and therefore limits the performance of the SWEDL model. Given the comparable performance of these two algorithms and considerations of simplicity, this study finally adopts the RIFTES framework without incorporating an explicit physical constraint in the loss function for LST retrieval.

## 6. Conclusion

LST is a fundamental parameter in TIR remote sensing. The TES and SW algorithms remain two most widely used methods for operational clear-sky LST retrieval, while both of them suffer from certain limitations. In recent years, hybrid methods combining the SW and TES algorithms have become increasingly popular. Nevertheless, these hybrid methods still face several challenges in general, including relatively high computational complexity caused by the iterative procedure inherited from TES, limited improvement in retrieval accuracy, and insufficient validation.

To address current research gaps, this study develops the RIFTES framework for accurate and efficient clear-sky LST retrieval, which consists of a lightweight deep residual neural network with masking mechanism for atmospheric correction and the non-iterative TES algorithm for separating LST and LSE. The neural network integrates physical mechanisms from atmospheric RTM and the SW method, enabling flexible implementation using readily available satellite observations with optional PWV input. The non-iterative TES algorithm is derived from physical principles to eliminate the iterative process of the original NEM module, reducing the computational complexity while retaining the physical foundation of the TES algorithm.

Comprehensive validations and comparisons demonstrate the effectiveness of the new algorithm. RIFTES generally achieves better retrieval accuracy in both the simulation analysis and T-based validation, maintaining robust across different sensors and conditions. Even when relying solely on remote sensing observations without PWV, the algorithm still achieves good performance. Furthermore, by simplifying both the atmospheric correction step and the NEM module, the RIFTES algorithm effectively reduces computational costs compared to the conventional TES and hybrid methods.

Therefore, by integrating the powerful learning capability of neural networks with multi-source physical foundations, the RIFTES algorithm not only achieves more accurate and robust LST estimates, but also effectively reduces the input requirements and computational costs. Such a dual-benefit characteristic highlights the promising potential of RIFTES for operational retrieval applications. The proposed clear-sky LST retrieval algorithm, together with our previously developed cloudy-sky LST retrieval method and spatial downscaling approach (Zhang et al., 2026; Zhang et al., 2024), will be applied in the near future to estimate all-sky LST at both high spatial and temporal resolutions.

## Acknowledgement

This work was supported in part by the National Natural Science Foundation of China under Grant 42230109, in part by the Yunnan International Joint Laboratory for Integrated Sky-Ground Intelligent Monitoring of Mountain Hazards under Grant 202403AP140002, in part by the Yunnan Plateau Remote Sensing Innovation Team under Grant 202505AS350001, and in part by the Yunling Scholar Project of the "Xingdian Talent Support Program" of Yunnan Province under Grant 41961053. We sincerely thank Dr. José A. Sobrino and Dr. Dražen Skoković for providing the in-situ LST measurements from the GCU network.